\documentclass[12pt]{iopart}
\usepackage{graphicx}
\usepackage{bm}
\usepackage{url}
\usepackage{hyperref}
\usepackage{mathrsfs}
\expandafter\let\csname equation*\endcsname\relax
\expandafter\let\csname endequation*\endcsname\relax
\usepackage{amsmath}
\usepackage{amssymb}
\usepackage{amsthm}
\graphicspath{{Figures/}}

\providecommand{\Hi}{\mathscr{H}}
\providecommand{\Uk}{U_{\mathrm{kick}}}

\providecommand{\abs}[1]{\left\lvert#1\right\rvert}
\providecommand{\Norm}[1]{\left \lVert #1 \right\rVert}
\newtheorem{thm}{Theorem}[]

\newtheorem{lemma}{Lemma}[]
\theoremstyle{definition}

\begin{document}

	\title{Generalized Product Formulas and  Quantum Control}
	
	\author{Daniel Burgarth$^{1}$, Paolo Facchi$^{2,3}$, Giovanni Gramegna$^{2,3}$, Saverio Pascazio$^{2,3,4}$}
	\address{$^{1}$Department of Physics and Astronomy, Macquarie University, NSW 2109,	Australia}
	\address{$^{2}$Dipartimento di Fisica , Universit\`a di Bari, I-70126 Bari, Italy}
	\address{$^{3}$INFN, Sezione di Bari, I-70126 Bari, Italy}
	\address{$^{4}$Istituto Nazionale di Ottica (INO-CNR), I-50125 Firenze, Italy}

	\begin{abstract}
		We study the quantum evolution under the combined action of the exponentials of two not necessarily commuting operators. We consider the limit in which the two evolutions alternate at infinite frequency. This case appears in a plethora of situations, both in physics (Feynman integral) and mathematics (product formulas).
		We focus on the case in which the two evolution times are scaled differently in the limit and generalize standard techniques and results.
	\end{abstract}
	
	\maketitle

	\section{Introduction}
	In most product formulas~\cite{ref:Trotter,ref:KatoTrotter,ref:Lie}, there is a subtle interplay between two competing dynamics. Such interplay has multiple facets, both physical (as for example in the quantum Zeno effect~\cite{ref:QZEMisraSudarshan,ref:QZEreview-JPA}) and mathematical~\cite{ref:kickfix}.
	In physics, the seminal ideas can be traced back to Feynman, who, working on his path-integral formulation of quantum mechanics 
	\cite{ref:Feynman(1948),ref:Dirac1933,ref:FeynmanHibbs(1965),Schulman1981}, 
	wrote the full dynamics of a quantum particle in the form 
	\begin{eqnarray}
	\rme^{-\rmi t H} = \rme^{-\rmi t (T+V)} = \lim_{n\rightarrow\infty} \bigl(\rme^{-\rmi \frac{t}{n} T} \rme^{-\rmi \frac{t}{n} V}\bigr)^n,
	\label{eq:feynmanTV}
	\end{eqnarray}
	where $H=T+V$ is the Hamiltonian, and $T$ and $V$ the kinetic and potential energy, respectively. 
	Feynman was attacking the formidable problem of calculating the exponential of the sum of two non-commuting operators.
	In mathematics, a similar problem was first posed by Lie~\cite{ref:Lie}, who proved that 
	\begin{eqnarray}
	\label{eq:lie}
	\rme^{A+B} = \lim_{n\rightarrow\infty} \bigl(\rme^{A/n} \rme^{B/n}\bigr)^n,
	\end{eqnarray}
	for square matrices $A$ and $B$. 
	
	Formulas~(\ref{eq:feynmanTV}) and~(\ref{eq:lie}) disguise, among serious mathematical difficulties, a subtle (and intriguing) standpoint: when one factors the exponentials, one always implicitly assumes that $n$ appears \emph{at the first power} in the denominator of the exponents. One is so accustomed to such a stance, that other scalings have not been looked at. What, then, about evolutions of the following type
	\begin{eqnarray}
	\label{eq:lie2}
	\bigl(\rme^{A/n^\gamma} \rme^{B/n}\bigr)^n ?
	\end{eqnarray}
	In the above formula, $\gamma$ is in general different from one or, alternatively, the evolution times under the action of the kinetic and potential energies in Eq.~(\ref{eq:feynmanTV}) are scaled differently. The most interesting situations arise when $0 \leq \gamma \leq 1$ (for $\gamma >1 $ the limit is trivially $\rme^{B}$, while for $\gamma <0$ the limit might not exist, as we will see later).
	In this Article we will investigate the mathematical features and limits of expressions of the type~(\ref{eq:lie2}). One expects that the factor $\rme^{A/n^\gamma}$ dominates over $\rme^{B/n}$ for $0 \leq \gamma < 1$, leading to quantum control (in the sense that $B$ will be \emph{modified} into an effective generator $B_Z$ yielding a controlled dynamics characterized by superselection sectors, as explained in section~\ref{sec:quantumcont}). We will indeed see that these formulas yield quantum Zeno subspaces~\cite{ref:QZS,ref:artzeno}, that are robust against the detrimental effects of decoherence. This observation provides a strong physical motivation for our analysis.
	
Our analysis will be organized as follows. In Sec.~\ref{sec:quantumcont} we  revisit two standard control techniques---frequently kicked evolution and strong continuous coupling---by exhibiting bounds on the control errors. We show that the two protocols only differ in the order a double limit is taken. Then, in Section~\ref{sec:intermediateLim}, we show that it is still possible to get quantum control in an intermediate situation, where the operators in the exponentials scale differently with~$n$. Finally, in Section~\ref{sec:GeneralizedPF}, as a byproduct of our results, we  discuss the generalization~\eqref{eq:lie2} of the Trotter product formula, by providing analytical  bounds on the convergence rate and by comparing them with a numerical analysis.  Four appendices are devoted to the proofs of the theorems.
	
	\section{Preliminaries: notation and quantum control}
	\label{sec:quantumcont}
	
	We shall first introduce notation by adhering to the terminology of quantum applications, and then look in detail at two different quantum control protocols, examining similarities and differences.
	
\begin{figure}[t]\centering
		\includegraphics[width=.5\textwidth]{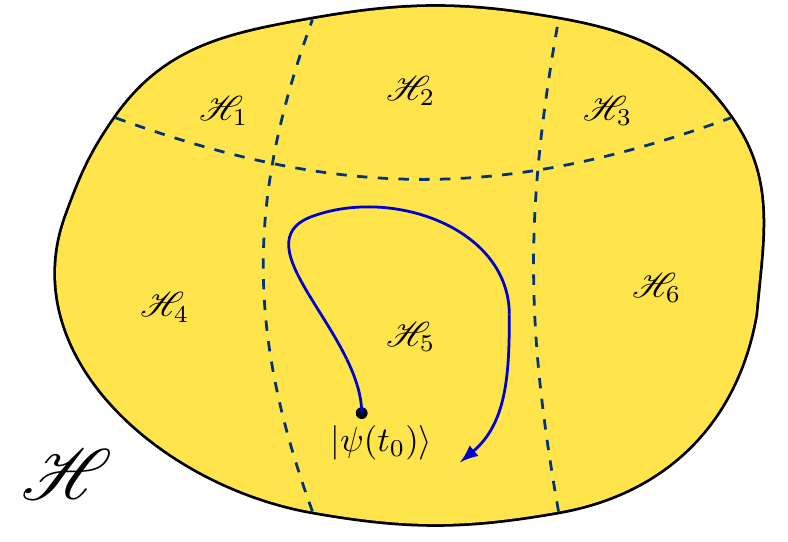}
		\caption{Pictorial representation. The Hilbert space $\Hi$ is partitioned into quantum Zeno subspaces $\Hi_\mu=P_\mu \Hi$. If the system is in a given  subspace (say $\Hi_5$) at the initial time $t_0$, it will coherently evolve within this subspace and will never make transitions to other subspaces.}
		\label{fig:HZ}
	\end{figure}

	Consider a quantum system living in a Hilbert space $\Hi$ with finite dimension, $\dim\Hi<\infty$. Let $U(t)=\rme^{-\rmi tH}$ be the (``free") evolution operator, $H$ being the Hamiltonian of the system. Let $\{P_\mu \}$ be a complete  family of orthogonal projections, that is a set of $m$ projection operators, with $m\leqslant \dim\Hi$, satisfying
	\begin{equation}
	P_\mu^\dagger=P_\mu,\qquad P_\mu P_\nu=\delta_{\mu\nu}P_\mu,\qquad \sum_{\mu=1}^m P_\mu=I.
	\label{eq:pmu}	
	\end{equation}
	The aim of quantum control, in the context of decoherence suppression, is to engineer an evolution in which the Hilbert space is dynamically partitioned 
	\begin{equation}
	\Hi=\bigoplus_{\mu=1}^m{\Hi_\mu} ,
	\end{equation} 
	so that transitions between different subspaces $\Hi_\mu = P_\mu \Hi$ are suppressed. See Fig.~\ref{fig:HZ}. The subspaces will be called quantum Zeno subspaces and the control procedures will be referred to as quantum Zeno dynamics (QZD).
	
	\subsection{Frequently pulsed evolution}
	QZD can be obtained by applying frequent and instantaneous unitary transformations to the evolving state of the system. The control procedure consists in alternating free evolutions of the system with instantaneous unitary ``kicks"
	\begin{equation}
	U_n(t)=\bigl(\Uk \rme^{-\rmi \frac{t}{n}H}\bigr)^n .
	\end{equation}
	The ensuing control techniques were first investigated in the 60's in relation to magnetic resonance~\cite{anderson,ernst,freeman,levitt} and are often referred to as ``bang-bang" dynamics~\cite{viola98} in the more recent quantum literature. For a good review, see Ref.~\cite{lidarrev}.
	Evolutions of this type are of tantamount importance in the study of quantum chaos~\cite{ref:CasatiChaos,ref:BerryChaos,ref:Gutzwiller,ref:qchaos}, although in that case the frequency is kept finite and $t$ scaled like $n$.
	
	Let 
	\begin{equation}
	\Uk=\sum_{\mu=1}^m \rme^{-\rmi \phi_\mu}P_\mu ,
	\label{eq:specdecUk}
	\end{equation}
	with $m\leqslant \dim\Hi$, be the spectral decomposition of $\Uk$, where
	$\{P_\mu\}$ is a complete family of projections~\eqref{eq:pmu} and 
	$\rme^{-\rmi \phi_\mu}\neq \rme^{-\rmi \phi_\nu}$ for $\mu\neq\nu$.

	In the $n\rightarrow\infty$ limit  (infinitely frequent pulses applied in a fixed time interval $(0,t)$), one obtains a QZD with Zeno subspaces defined by the eigenspaces   of the unitary kick~(\ref{eq:specdecUk}). This is a consequence of the following
	\begin{thm}
		\label{thm:PulsedFormulation}
		Let $H$ be a Hermitian operator and $\Uk$ be a unitary operator on a finite dimensional Hilbert space $\Hi$. Then the following limit holds
		\begin{equation}\label{eq:PulsedLimit}
		\Uk^{\dagger n} U_n(t) \to \rme^{-\rmi tH_Z}, \qquad \text{as } n\to\infty,
		\end{equation} 
		where
		\begin{equation}\label{eq:HZ}
		H_Z=\sum_{\mu=1}^m P_\mu H P_\mu
		\end{equation}
		is the Zeno Hamiltonian with respect to the eigenprojections $\{P_\mu\}$ of $\Uk$. In particular, for large $n$ we get
		\begin{equation}
		\label{eq:UZK}
		U_n(t)=\Uk^{n}  \rme^{-\rmi tH_Z}+\mathcal{O}\left(\frac{1}{n}\right) .
		\end{equation}
	\end{thm}
	The proof is given in \ref{sec:app1}.
	As one can see, there is an important contribution of the Hamiltonian $H$ to the evolution, which stems from its diagonal part with respect to the unitary kick (note that $[\Uk,H_Z]=0$).  It is useful to re-write the above evolution as follows
	\begin{equation}
	U_n(t)=\rme^{-\rmi \sum_\mu (n\phi_\mu P_\mu+ t P_\mu H P_\mu)}+\mathcal{O}\left(\frac{1}{n}\right) 
	=\sum_{\mu=1}^m \rme^{-\rmi n\phi_\mu - \rmi t P_\mu H P_\mu}P_\mu+\mathcal{O}\left(\frac{1}{n}\right).
	\label{eq:un}
	\end{equation}
	This expression clarifies that the system evolves in each subspace $\Hi_\mu = P_\mu \Hi$ of the kick operator according to the projected Hamiltonian $P_\mu H P_\mu$, with a subspace-dependent phase $n\phi_\mu$.

	\subsection{Strong Continuous Coupling}
	QZD can also be obtained by coupling the system with Hamiltonian $H$ to a (control) potential $V$. The evolution is
	\begin{equation}
	U_K(t)=\rme^{-\rmi t(H+KV)},
	\end{equation} 
	where $K$ is the coupling constant, to be taken large if one aims at getting a good control procedure.
	
	Let 
	\begin{equation}\label{eq:Vspecdec}
	V=\sum_{\mu=1}^m \lambda_\mu P_\mu
	\end{equation}
	be the spectral decomposition of the control potential $V$,
	where $\lambda_\mu$'s are the (possibly degenerate) distinct eigenvalues of $V$ and $P_\mu$'s the corresponding eigenprojections, satisfying conditions 
	(\ref{eq:pmu}).

	Naively, one might expect that, as $K\rightarrow\infty$, it becomes possible to neglect the action of $H$, so that the system evolves under the sole action of the control potential $V$. Transition among different subspaces would be avoided and the state would simply acquire a subspace-dependent phase. However, a more careful analysis shows that Hamiltonian $H$ yields a non-trivial contribution to the limiting evolution. This is the consequence of the following
	\begin{thm}
		\label{thm:ContFormulation}
		Let $H$ and $V$ be Hermitian operators acting on a finite dimensional Hilbert space $\Hi$, with $V$ having the spectral decomposition~\eqref{eq:Vspecdec}. Then the following limit holds
		\begin{equation}\label{eq:StrCoupLim}
		\rme^{\rmi tKV}\rme^{-\rmi t(H+KV)}\to \rme^{-\rmi tH_Z} , \qquad \text{as } K\to\infty, 
		\end{equation}
		uniformly on compact time intervals, where $H_Z$ is the Zeno Hamiltonian~\eqref{eq:HZ} with respect to the eigenprojections $\{P_\mu\}$ of $V$.

		In particular, for large $K$ we have
		\begin{equation}
		\label{eq:strcoupapproachrate}
		\rme^{-\rmi t(H+KV)}=\rme^{-\rmi tKV}\rme^{-\rmi tH_Z}+\mathcal{O}\left(\frac{1}{K}\right).
		\end{equation}
	\end{thm}
	The theorem is proved by going to the $H$-interaction picture and by using Kato's adiabatic theorem~\cite{ref:KatoAdiabatic,ref:avron,ref:unity1}. A rapid review of the adiabatic theorem is provided for completeness in~\ref{sec:appadiab}, while Theorem~\ref{thm:ContFormulation} is proved in~\ref{app:StrongCouplingProof}.
	
	As with a unitary kick, the contribution of the Hamiltonian $H$ to the limiting evolution stems from its diagonal part with respect to the control potential (note that $[V,H_Z]=0$).  Using the spectral decomposition of $V$ in Eq.~\eqref{eq:Vspecdec} the evolution operator can be written
	\begin{align}\notag
	U_K(t)&=\rme^{-\rmi t\sum_\mu K\lambda_\mu P_\mu+P_\mu H P_\mu}+\mathcal{O}\left(\frac{1}{K}\right) \\
	&=\sum_{\mu=1}^m \rme^{-\rmi t(K\lambda_\mu+P_\mu H P_\mu)}P_\mu+\mathcal{O}\left(\frac{1}{K}\right).
	\label{eq:uk}
	\end{align}
	Off-diagonal transitions (with respect to the eigenspaces of $V$) are suppressed, while in each $V$-eigenspace $\Hi_\mu$ the system evolves non-trivially according to the projected Hamiltonian $P_\mu H P_\mu$.

	\subsection{Similarities and differences}
	
	The limiting dynamics~(\ref{eq:un}) and~(\ref{eq:uk}) are strikingly similar. We now show that this similarity is a consequence of a double limit~\cite{ref:BBZeno}, where the order in which the two limits are taken is immaterial.
	
	Let us first observe that, using the Trotter product formula, one can get a continuous coupling starting from a pulsed-like evolution:
	\begin{equation}\label{eq:Trotter}
	\left(\rme^{-\rmi \frac{t}{n}KV}\rme^{-\rmi \frac{t}{n}H}\right)^n=\rme^{-\rmi t(H+KV)}+\mathcal{O}\left(\frac{1}{n}\right).
	\end{equation}
	Define the evolution operator
	\begin{equation}
	\label{eq:UnK}
	U_{n,K}(t)=\left(\rme^{-\rmi \frac{t}{n}KV}\rme^{-\rmi \frac{t}{n}H}\right)^n .
	\end{equation}
	The strong coupling limit~\eqref{eq:StrCoupLim} can be written 
	\begin{equation}
	\lim_{K\to\infty}\lim_{n\to\infty}\rme^{\rmi tKV}U_{n,K}(t)=\rme^{-\rmi tH_Z} ,
	\end{equation}
	where the inner $n$-limit yields continuous coupling, while the outer $K$-limit yields the strong coupling limit. 
	On the other hand, one gets the kicked (bang-bang) control~\eqref{eq:PulsedLimit} with $\Uk=\rme^{-\rmi tV}$ as
	\begin{equation}\label{standard}
	\lim_{K=n \to \infty}\rme^{\rmi tKV}U_{n,K}(t)=\rme^{-\rmi tH_Z} .
	\end{equation}
	Both cases make use of a double limit in the variables $n,K$. In the former case, the limit is first taken on $n$ and then on $K$, while in the latter case the limit is taken along the diagonal of the $(n,K)$ plane, as shown in Fig.~\ref{fig:(n,K)plane} (solid-blue line). The dashed-red line in Fig.~\ref{fig:(n,K)plane} corresponds to the limit $n\rightarrow\infty$, yielding the Trotter product formula~\eqref{eq:Trotter}, namely continuous coupling without the strong coupling limit. 
	
	A few warnings are necessary. Although the \emph{limiting} procedures are equivalent, the details and speed of convergence depend on (physical) procedures and experimental implementation~\cite{ref:berry,ref:ControlDecoZeno,Ketterle}, in particular because for $n$ or $K$ large but finite one might incur in the inverse Zeno effect~\cite{ref:AZE,ref:InverseZeno,ref:koshinoshimuzu}, whereby transitions to the other Zeno subspaces are accelerated, rather than suppressed~\cite{Wilkinson}. This is a crucial factor, in the light of the many recent experiments on QZD~\cite{Raimond2010,Firenze2014,Signoles2014,Wineland,Reichel}.
	
	\begin{figure}[t]
		\begin{center}
			\includegraphics[width=.55\textwidth]{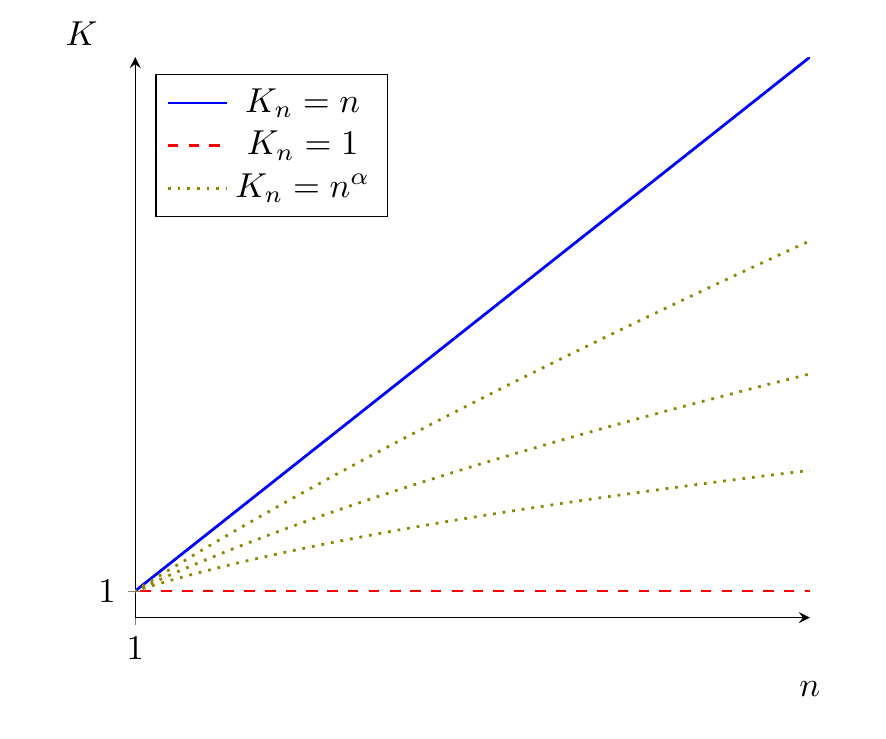}
		\end{center}
		\caption{The full (blue) line corresponds to the ``simultaneous" limit in $K$ and $n$, representing the pulsed dynamics. The dashed (red) curve corresponds to the limit taken only over $n$, which is the Trotter limit, yielding continuous coupling. In such a case there is no control since there is no strong coupling limit. The other curves refer to a coupling constant $K_n =n^\alpha$ with $0<\alpha<1$.}
		\label{fig:(n,K)plane}
	\end{figure}

	\section{An intermediate limit}\label{sec:intermediateLim}
	
	Motivated by the preceding comments, and by the pictorial view in Fig.~\ref{fig:(n,K)plane}, we now consider intermediate situations, and ask whether any interesting limit can be proved (therefore yielding quantum control) also in the region of the $(n,K)$ plane between the two extremal cases considered. The answer is affermative, and,
	in particular, the double limit along the curves 
	\begin{equation}
	K_n =n^\alpha, \qquad  \text{with } \alpha\in (0,1)
	\label{eq:Knnalpha}
	\end{equation} 
	yields quantum control, as shown in the following theorem.
	\begin{thm}\label{thm:doublelimit}
		Let $U_{n,K}(t)$ be the pulsed evolution~\eqref{eq:UnK},
		with $V$ having the spectral decomposition~\eqref{eq:Vspecdec}, and assume that 
		\begin{equation}
		K_n\to \infty, \qquad \text{with } K_n =o\left( n \right), \qquad  \text {as } n\rightarrow\infty. 
		\label{eq:Kncond}
		\end{equation}
		Then one has
		\begin{equation}
		\label{eq:IntermediateLimit}
		U_{n,K_n }(t)=\rme^{-\rmi tK_n V} \rme^{-\rmi tH_Z}+\mathcal{O}\left(\frac{1}{K_n }\right),
		\end{equation}
		as $n\to\infty$, uniformly on compact time intervals,
		where $H_Z$ is the Zeno Hamiltonian~\eqref{eq:HZ} with respect to the eigenprojections $\{P_\mu\}$ of $V$.
	\end{thm}

	The theorem is proved in \ref{sec:appinterm}. The error estimate in the above formula is obtained by using the same technique adopted for the pulsed procedure. The proof is a corollary of (the proof) of Theorem~\ref{thm:PulsedFormulation} when the unitary kick is given by
	\begin{equation}
	\Uk = \rme^{-\rmi \frac{t}{n}K_n V}, 
	\end{equation}
	whose spectral resolution is~\eqref{eq:specdecUk} with $\phi_\mu = t \lambda_\mu K_n/ n$. 
	Notice that, since by assumption $K_n/n\to 0$, for sufficiently large $n$ one has  $\max_{\mu,\nu} |\phi_\mu-\phi_\nu|\in (0,2\pi)$, 
	whence $\rme^{-\rmi \phi_\mu} \neq \rme^{-\rmi \phi_\nu}$ for all~$\mu\neq\nu$, and the eigenprojections of $V$ and $\Uk$ coincide.
		
\begin{figure}[t]
		\begin{center}
			\includegraphics[width=\textwidth]{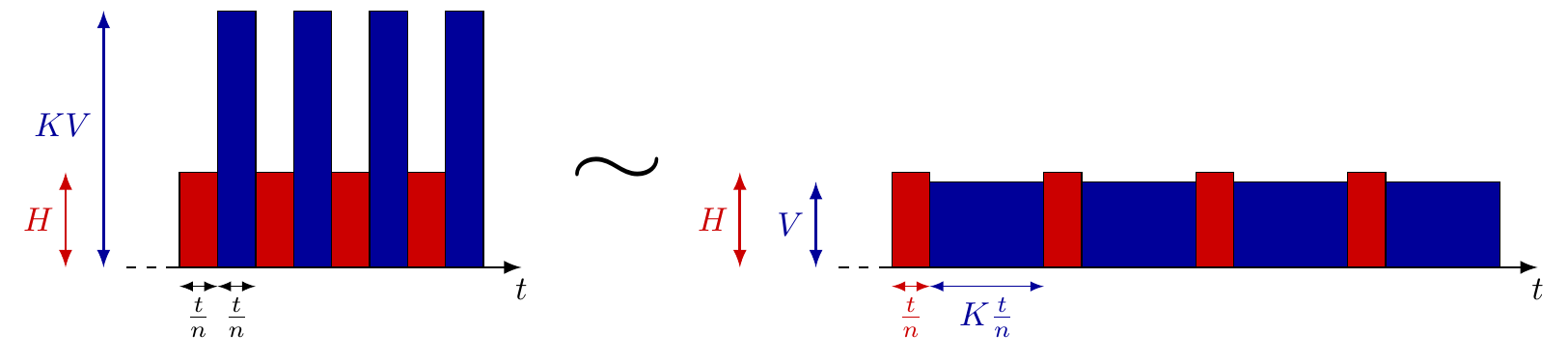}
		\end{center}
		\caption{Two equivalent ways of viewing the Trotter dynamics.}
		\label{fig:picfeynman}
	\end{figure}

	We now pause for a moment and give a pictorial view of the evolution~(\ref{eq:UnK}), yielding the limit~(\ref{eq:IntermediateLimit}). This can be interpreted in two equivalent ways. One can assume that each unitary acts for a time $t/n$, but the two generators $H$ and $KV$ are scaled differently, with $K=K_n$ as in Eq.~(\ref{eq:Kncond}): see left panel in Fig.~\ref{fig:picfeynman}. Alternatively, one can consider the two generators $H$ and $V$ acting for different times $t/n$ and 
	$Kt/n$, with $K=K_n$ as in Eq.~(\ref{eq:Knnalpha}): see right panel in Fig.~\ref{fig:picfeynman}. The two pictures are equivalent, and in both cases the factor $\rme^{-\rmi tK_n V}$ ``dominates" (controls) $\rme^{-\rmi tH}$ in~(\ref{eq:UnK}), the latter yielding the controlled dynamics $\rme^{-\rmi tH_Z}$ in~(\ref{eq:IntermediateLimit}), that acts within the Zeno subspaces.

	Applying Theorem~\ref{thm:doublelimit} to a coupling $K_n$ of the form~\eqref{eq:Knnalpha} we get
	\begin{equation}\label{eq:IntermediateLimitalpha}
	\bigl(\rme^{-\rmi t \frac{n^\alpha}{n}  V}\rme^{-\rmi \frac{t}{n}H}\bigr)^n
	=\rme^{-\rmi t n^\alpha V} \rme^{-\rmi tH_Z}+\mathcal{O}\left(\frac{1}{n^\alpha }\right),
	\end{equation}
	for all $\alpha \in (0,1)$. This represents a first step towards our seminal motivation, Eq.~(\ref{eq:lie2}). 
	
	Notice that the control can be extended up to  $\alpha=1$, i.e.\ $K_n=n$, for all non-resonant times $t$ such that 
\begin{equation}
\rme^{-\rmi t \lambda_\mu} \neq \rme^{-\rmi t\lambda_\nu}, \qquad  \text{for all } \mu\neq\nu
\label{eq:nonres}
\end{equation} 
(this requirement is needed in order to assure that the eigenprojections of $\rme^{-\rmi tK_n V}$ and $V$ are the same). 
	However, in general, for $\alpha>1$ the limit may not exist due to resonances: assume for example that $V^2=I$, then
	\begin{equation}
	U_{n,n^2}(\pi/2)=\left(\rme^{-\rmi n \frac{\pi}{2} V}\rme^{-\rmi \frac{\pi}{2 n}H}\right)^n
	\end{equation}
	does not have a limit. Indeed, for even $n$, $\rme^{-\rmi n \frac{\pi}{2} V}=(-1)^{n/2} I$, and thus the control is ineffective, $U_{n,n^2}(\pi/2)=\left(\rme^{-\rmi \frac{\pi}{2 n}H}\right)^n = \rme^{-\rmi \frac{\pi}{2} H}$, while for odd $n$, Eq.~\eqref{eq:IntermediateLimit} holds.
	
	Similar phenomena  were obtained for $\alpha>1$ in the context of the quantum Zeno effect, where  the limit evolution was shown to be sensitive to the spectral properties of the periodic projections and to the arithmetic properties of~$\alpha$~\cite{Zenolimit}.

\begin{figure}[t]
		\centering 
		\includegraphics[height=11.9 em]{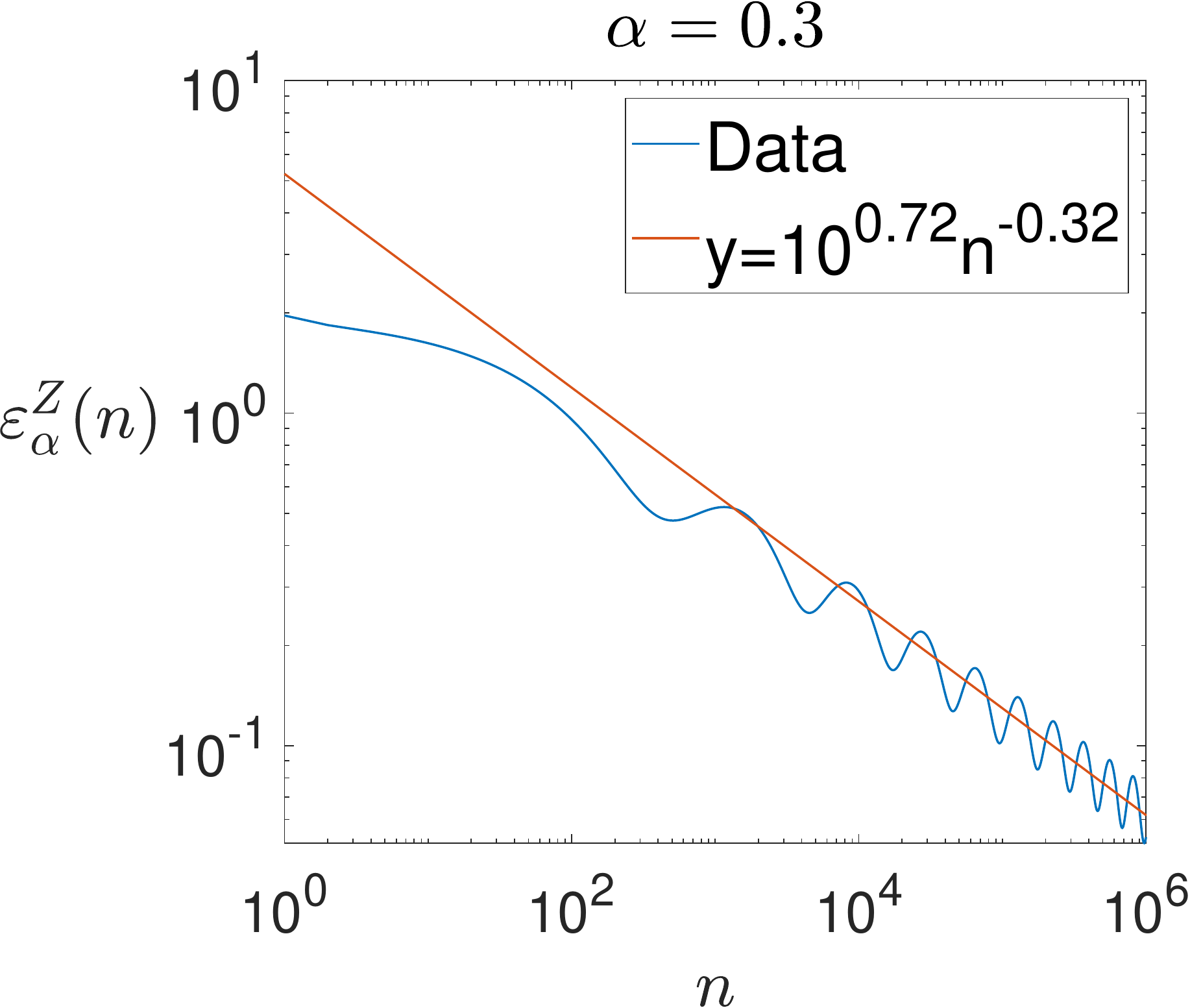}%\vspace{5pt}
		\includegraphics[height=11.9 em]{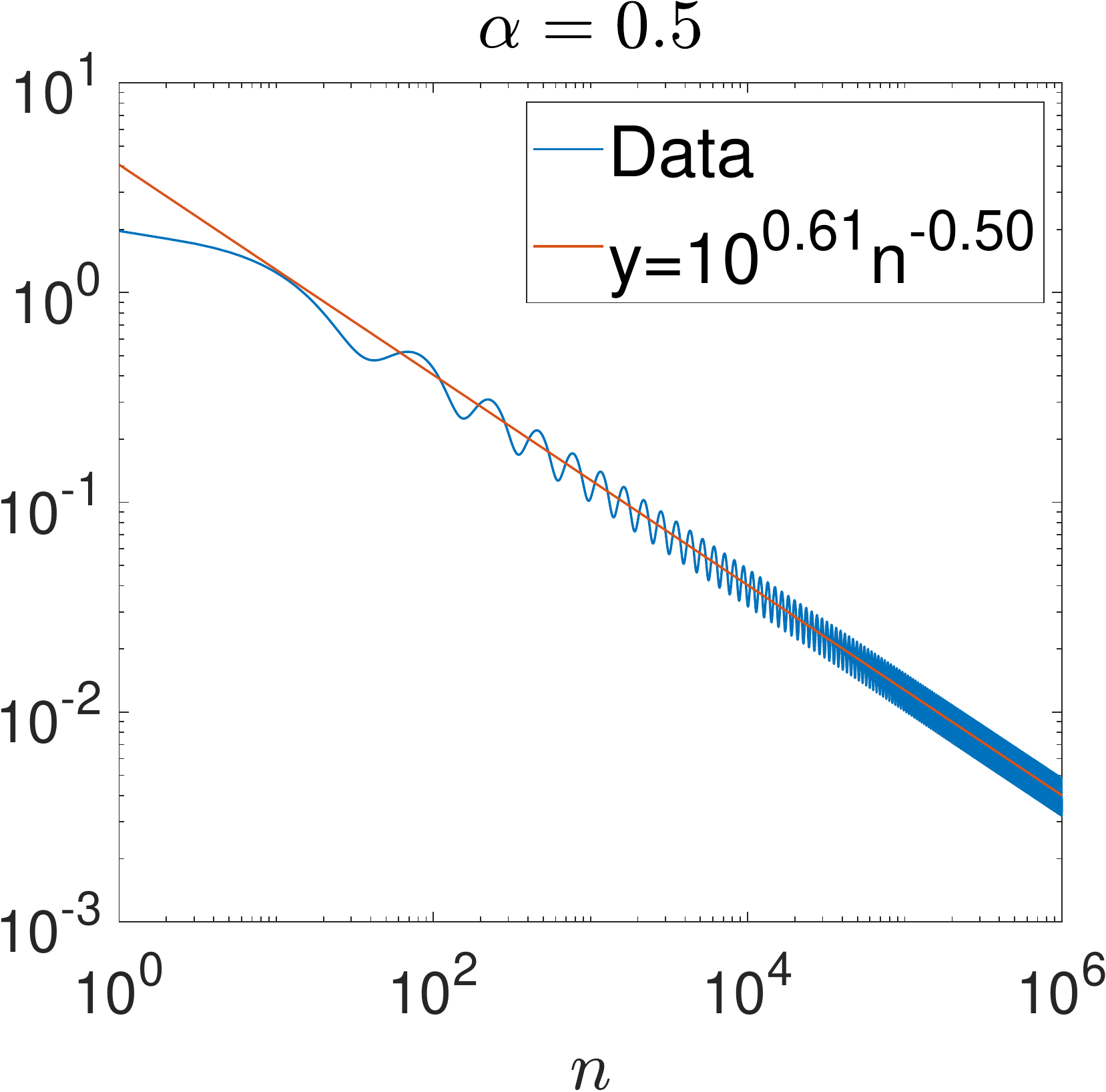}%\vspace{5pt}
		\includegraphics[height=11.9 em]{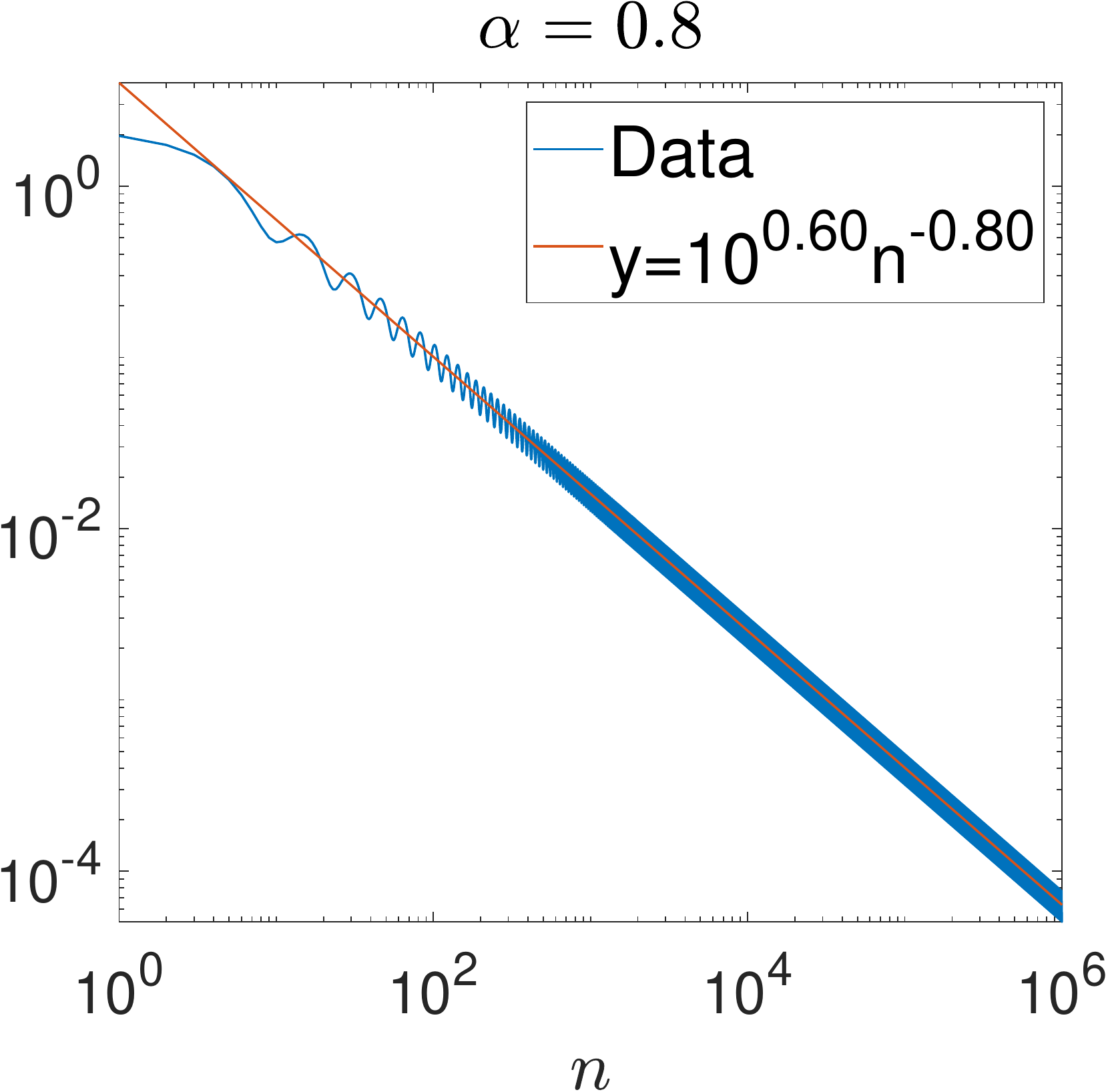}
		\caption{Error~(\ref{eq:IntermediateLimit}), as defined in Eq.~(\ref{eq:errornum}), for three different values of $\alpha$. The fit always 
		yields an error $\mathcal{O}(K_n^{-1})=\mathcal{O}(n^{-\alpha})$}
		\label{fig:ZenoAlpha}
	\end{figure}

	A numerical analysis shows that the error estimate in~\eqref{eq:IntermediateLimitalpha} is indeed sharp. See Fig.~\ref{fig:ZenoAlpha}.
	We perform the numerical simulation by considering $5\times 5$ matrices and set $t=1$. For the free Hamiltonian $H$, we generate a  matrix $A$ with random entries in the square $[-1,1]\times[-i,i]$ of the complex plane, and consider the Hermitian matrix
	\begin{equation}\label{eq:randomHermitian}
	H=\frac{A+A^\dagger}{2}.
	\end{equation}
	For the control potential $V$, we take a matrix with two eigenspaces, a $2$- and a $3$-dimensional one: $V=\mathrm{diag}(\lambda_1,\lambda_1,\lambda_2,\lambda_2,\lambda_2)$. The particular choice of $\lambda_1,\lambda_2$ is irrelevant as long as they are different. We set $\lambda_1=1,\lambda_2=0$.
	Let
	\begin{equation}
	\label{eq:errornum}
	\varepsilon_\alpha^Z(n)=\bigl\| U_{n,n^\alpha}(t)- \rme^{-\rmi t n^\alpha V} \rme^{-\rmi tH_Z} \bigr\| ,
	\end{equation}
	where $\Norm{A}:=\sqrt{\tr{(A^\dagger A)}}$ is the Hilbert-Schmidt norm. 
	In order to determine the asymptotic behaviour, we take a linear fit of the above quantity over the last decade of points in a logarithmic plot. Figure~\ref{fig:ZenoAlpha} displays our results. One observes that for three different values of $\alpha$, the distance~(\ref{eq:errornum}) decays like $K_n^{-1}=n^{-\alpha}$, proving that the limit 
	\eqref{eq:IntermediateLimit} is sharp.

	\section{Generalized product formula}\label{sec:GeneralizedPF}
	
	The link between the pulsed dynamics and the continuous coupling has been established using the Trotter approximation~\eqref{eq:Trotter}, where the two parameters $n$ and $K$ are considered independent. By contrast, in the intermediate situation considered in the previous section, these parameters satisfy a given relation $K=K_n $. Loosely speaking, a glance at Eq.~(\ref{eq:IntermediateLimit}) suggests that one manages to control the dynamics of the system as if the Trotter product formula were valid, despite the dependence $K=K_n $. To see this, note that by comparing the asymptotics~\eqref{eq:IntermediateLimit}, 
	\begin{equation}\label{eq:intermediateapprorate}
	\bigl(\rme^{-\rmi \frac{t}{n}K_n V}\rme^{-\rmi \frac{t}{n}H}\bigr)^n=\rme^{-\rmi tK_n V} \rme^{-\rmi tH_Z}+\mathcal{O}\left(\frac{1}{K_n }\right),
	\end{equation}
	with the strong coupling limit~\eqref{eq:strcoupapproachrate},
	\begin{equation}\label{eq:strcouplapprorate2}
	\rme^{-\rmi t(H+K_n  V)}=\rme^{-\rmi tK_n V} \rme^{-\rmi tH_Z}+\mathcal{O}\left(\frac{1}{K_n }\right),
	\end{equation}
	one gets 
	\begin{equation}\label{eq:GeneralizedAnalyticBound2}
	\bigl(\rme^{-\rmi \frac{t}{n}K_n V}\rme^{-\rmi \frac{t}{n}H}\bigr)^n=\rme^{-\rmi t(H+K_n  V)}+\mathcal{O}\left(\frac{1}{K_n }\right),
	\end{equation}
	as $n\to\infty$, with $K_n $ satisfying~\eqref{eq:Kncond}.
	
	This equation resembles the Trotter product formula, except for the $n$-dependence of the coupling constant $K_n $, suggesting that an approximation of this sort might be valid in more general situations. Due to many physical applications, the extended validity of Trotter's formula is interesting in its own right, so that it would be desirable to understand under which conditions this approximation can be used and which errors are implied. 
	One gets the following result.
	\begin{thm}
		\label{thm:GeneralizedProductFormula}
		Let $H$ and $V$ be Hermitian operators acting on a finite dimensional Hilbert space $\Hi$ and let $K_n $ be a real-valued function of $n$ such that  
		\begin{equation}
		K_n =o\left( n \right), \qquad  \text {as } n\rightarrow\infty. 
		\end{equation}
		Then
		\begin{equation}
		\label{eq:GeneralizedProductFormula2}
		\lim_{n\rightarrow\infty}\left[\bigl(\rme^{-\rmi \frac{t}{n}K_n V}\rme^{-\rmi \frac{t}{n}H}\bigr)^n-\rme^{-\rmi t(K_n +H)}\right]=0.
		\end{equation}
		In particular, for large $n$ one has
		\begin{equation}\label{eq:GeneralizedAnalyticBound}
		\bigl(\rme^{-\rmi \frac{t}{n} {K_n }V}\rme^{-\rmi \frac{t}{n}H}\bigr)^n-\rme^{-\rmi t( {K_n } V +H)}=\mathcal{O}\left(\frac{K_n }{n}\right).
		\end{equation}
	\end{thm}
	This is proved in~\ref{app:GenTrotter}, by exploiting the usual techniques adopted to prove the classical Trotter product formula.
	
	By comparing~\eqref{eq:GeneralizedAnalyticBound} with~\eqref{eq:GeneralizedAnalyticBound2} we see immediately that the error bound in Theorem~\eqref{thm:GeneralizedProductFormula} is not optimal, since for  $K_n =n^\alpha$ with $1/2<\alpha<1$, Eq.~\eqref{eq:GeneralizedAnalyticBound2} establishes a better bound. Using these two estimates together we can establish that the error bound is smaller than $\mathcal{O}(1/\sqrt{n})$ and the worst case occurs at $\alpha=1/2$. 
	
	\subsection{Numerical analysis}
	\label{numan}
	
	\begin{figure}[t]
		\centering
		\includegraphics[height=11 em]{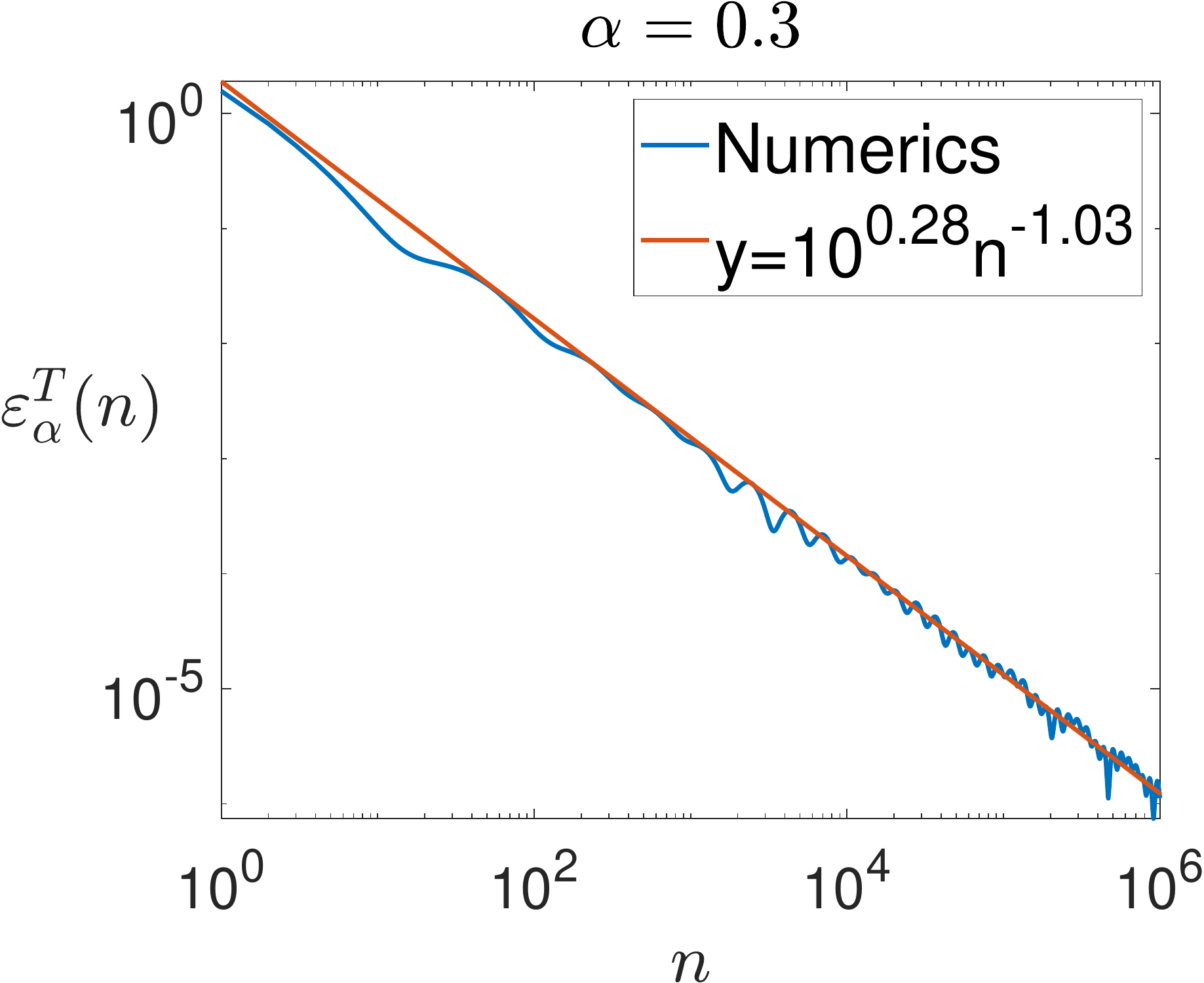}%\vspace{5pt}
		\includegraphics[height=11 em]{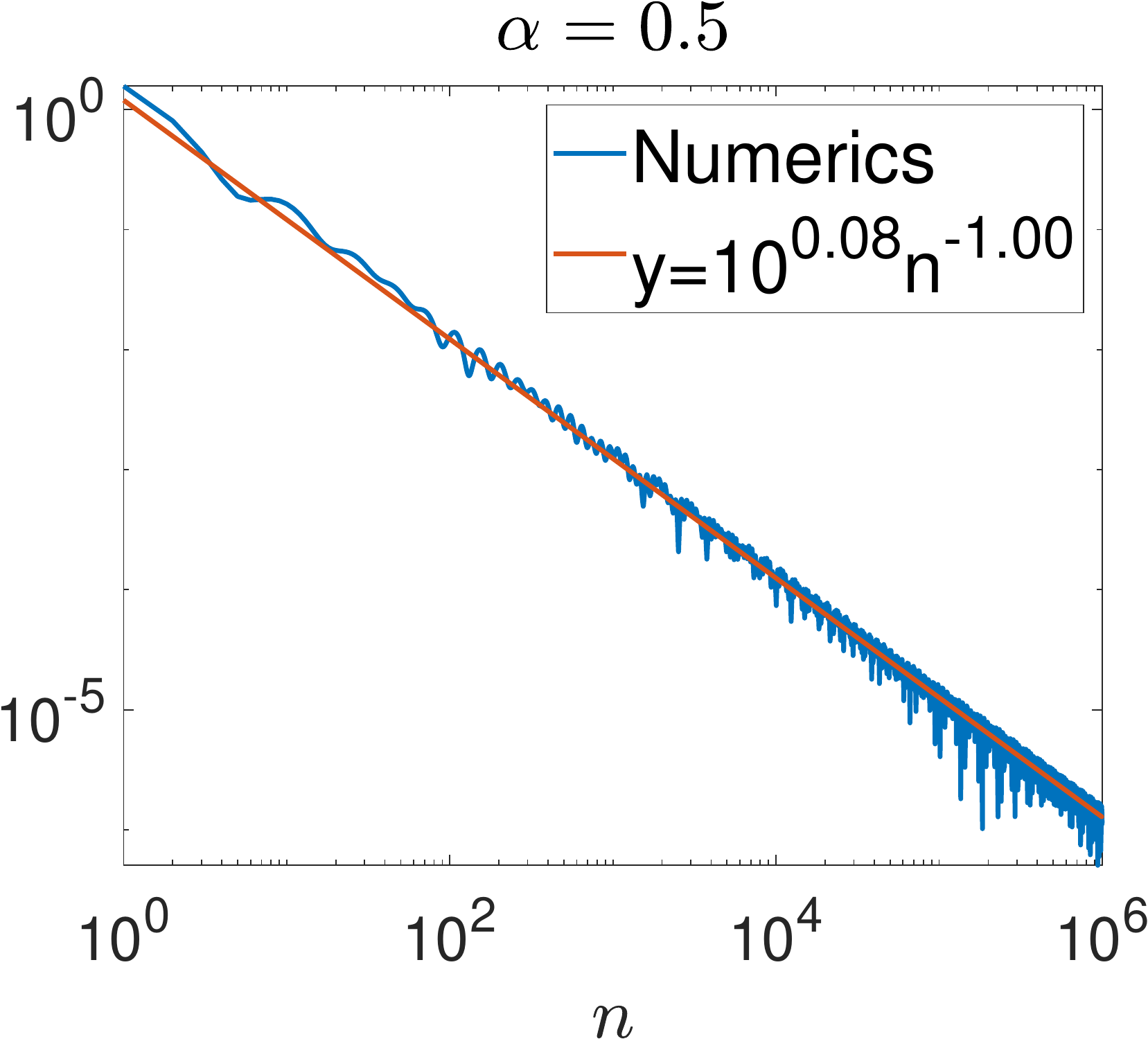}%\vspace{5pt}
		\includegraphics[height=11 em]{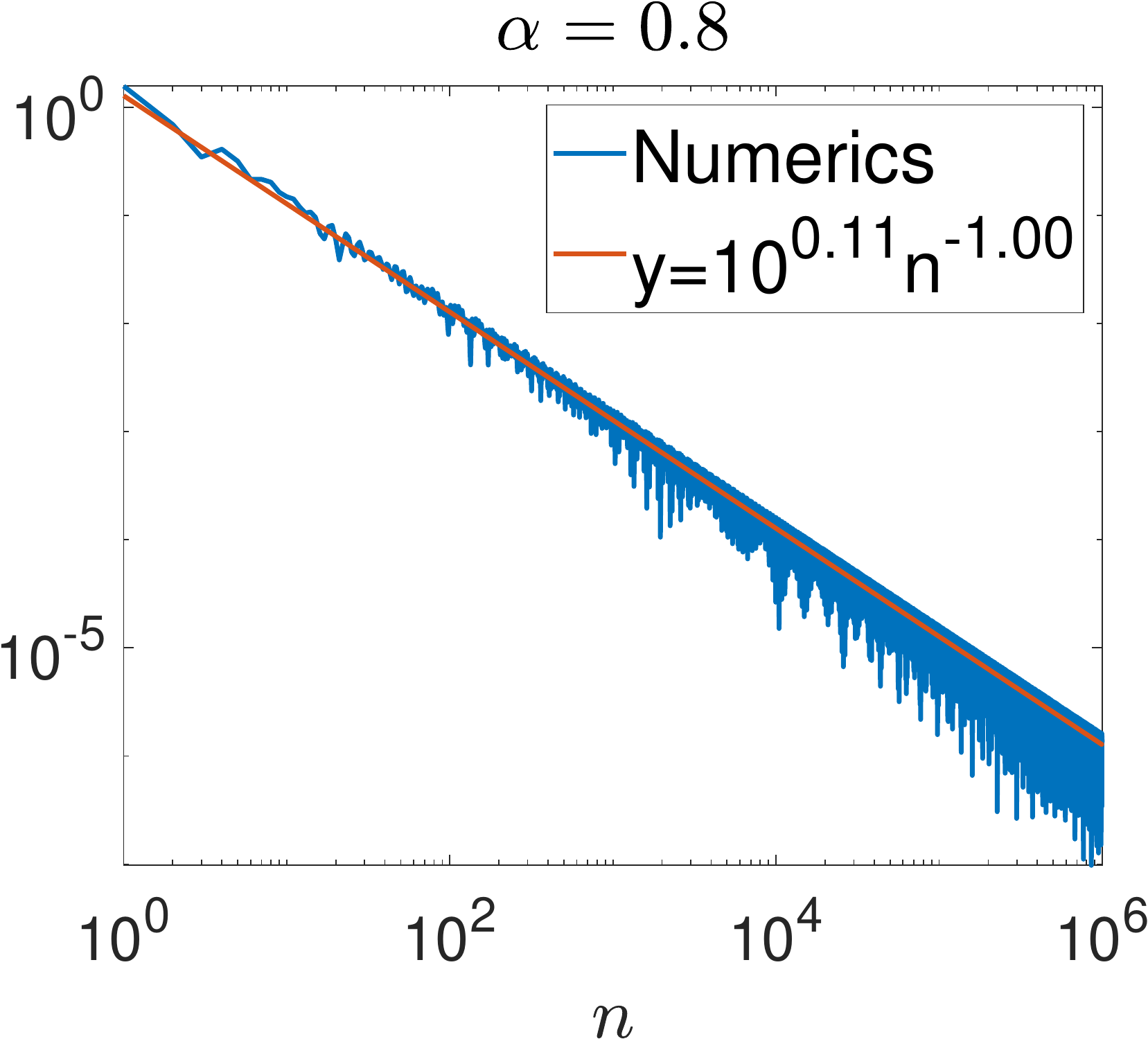}
		\caption{The error~(\ref{eq:errt}) is independent of $\alpha$ and is always $\mathcal{O}(1/n)$.
			}
		\label{fig:TrotterAlpha}
	\end{figure}

	We have performed a numerical analysis of the generalized product formula~(\ref{eq:GeneralizedProductFormula2}), using random Hermitian matrices $H$ and $V$ generated as in~\eqref{eq:randomHermitian} and analyzing the quantity 
	\begin{equation}
	\label{eq:errt}
	\varepsilon_\alpha^T(n)=\bigl\| \bigl(\rme^{-\rmi \frac{t}{n} n^\alpha V}\rme^{-\rmi \frac{t}{n}H}\bigr)^n-\rme^{-\rmi t(n^\alpha V+H)}\bigr\|
	\end{equation}
	as a function of $n$. The results are displayed in Fig.~\ref{fig:TrotterAlpha}, and are of interest, in that they show that the error is always $\mathcal{O}(n^{-1})$.  We offer no analytic explanation, at this stage, for this bound.
	
\begin{figure}[t]
		\centering
		\includegraphics[width=.45\textwidth]{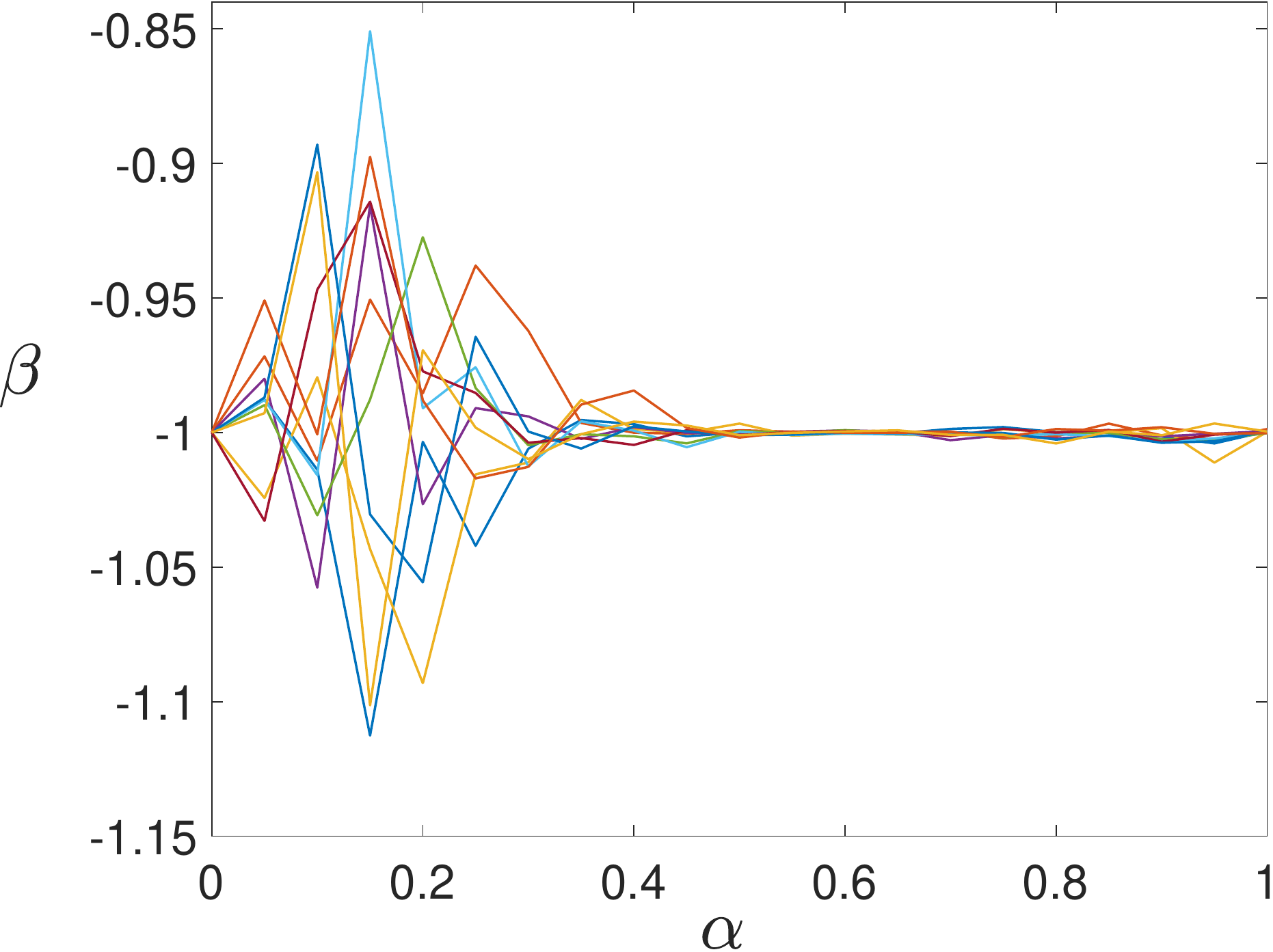}
		\includegraphics[width=.45\textwidth]{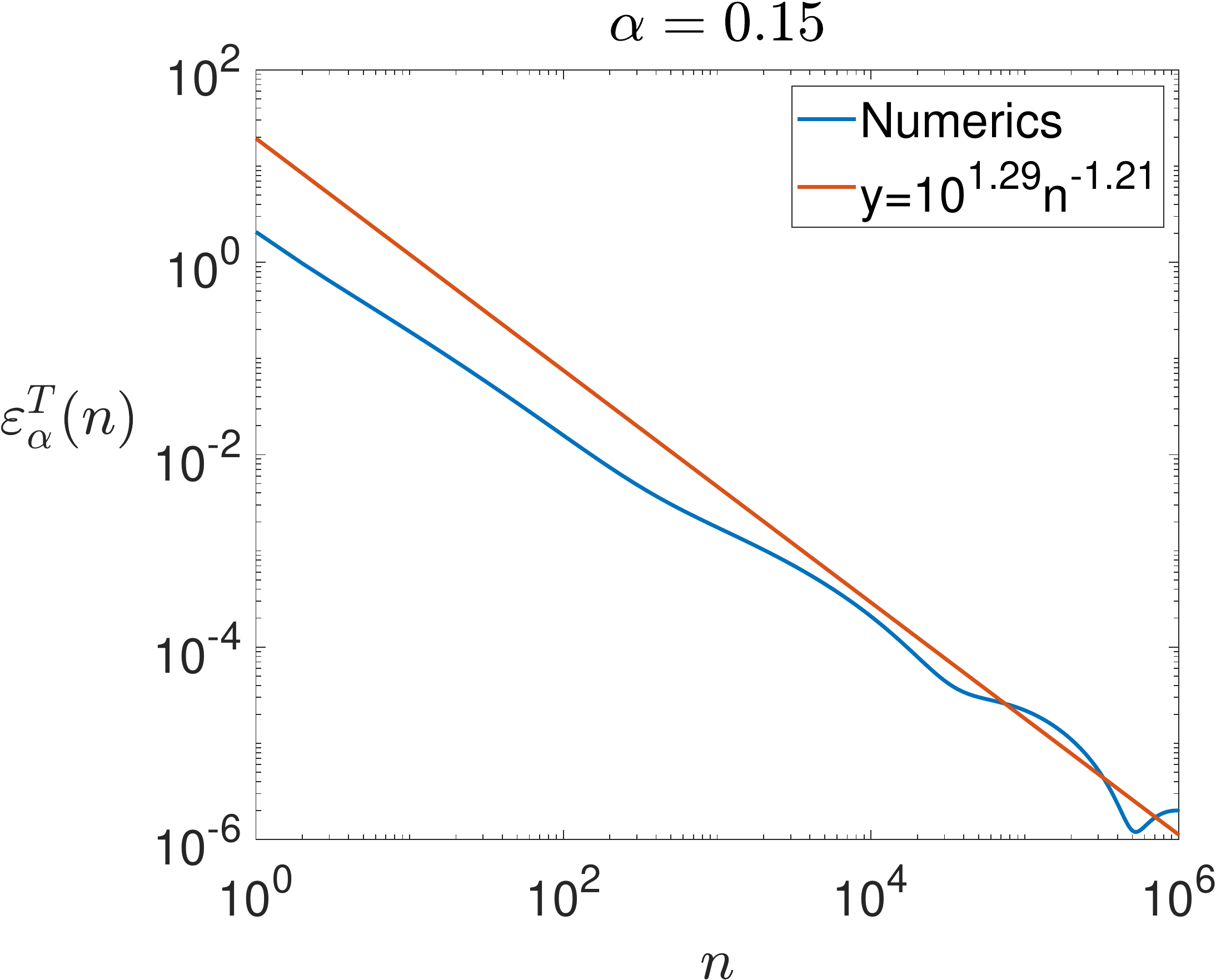}
		\caption{\textit{Left panel}: Exponent $\beta$ of the asymptotic power-like behavior $\varepsilon^T_\alpha(n)=\mathcal{O}(n^\beta)$ vs $\alpha$.  Different curves correspond to different random iterations. The exponent is essentially $-1$ for large enough $\alpha$.
		\textit{Right panel}: Error $\varepsilon^T_{\alpha=0.15}(n)$, in the region of the left panel
			where the power-like behaviour is less stable. The region where the linear fit is computed, between $n=10^4$ and $n=10^6$, is not in the full asymptotic regime and displays oscillations that distort the linear fit. }
		\label{fig:TrotterAllAlpha}
	\end{figure}

	In order to better characterize the asymptotic behaviour for different values of $\alpha\in(0,1)$, we performed further numerical analyses in accordance with the following procedure:
	i) divide the interval $[0,1]$ in equal steps $\Delta\alpha=0.05$ (yielding $21$ values for $\alpha$); 
	ii) for each value of $\alpha$, perform a linear fit of the curve $\varepsilon_\alpha^T(n)$ in a logarithmic scale for two decades of points between $n=10^4$ and $n=10^6$;
	iii) plot the exponent $\beta$ of the asymptotic power behaviour of $\varepsilon_\alpha^T(n)$ vs $\alpha$;
	iv) iterate the procedure for several random matrices ($N_{\text{iter}}=10$).
	The results obtained from this procedure are shown in Fig.~\ref{fig:TrotterAllAlpha} (different curves corresponding to different random iterations) and confirm that the power behaviour of $\varepsilon_\alpha^T(n)$ is independent of $\alpha$, and yields with very good approximation
	\begin{equation}
	\label{eq:numerr}
	\left(\rme^{-\rmi \frac{t}{n}K_n V}\rme^{-\rmi \frac{t}{n}H}\right)^n=\rme^{-\rmi t(K_n V+H)}+\mathcal{O}\left(\frac{1}{n}\right).
	\end{equation}
	The numerical analysis confirms that the Trotter product formula works exactly as if $K$ were independent of $n$. 
	
	We conclude this section with a final comment on the oscillations ($\simeq 10\%$) of the exponent around the value $\alpha\simeq 0.15$, that are observed in Fig.~\ref{fig:TrotterAllAlpha}. They can be explained by scrutinizing a particular iteration about this value, see the right panel of  Fig.~\ref{fig:TrotterAllAlpha}. 
	The evolution has not reached yet the asymptotic regime for $n=10^4-10^6$, and oscillations distort the linear fit.
	
	\subsection{The qubit case}
	\label{sec:qubit}
	We corroborate the independence of the convergence rate in~\eqref{eq:numerr} from $\alpha$, as
	shown by the numerics, by providing an explicit example for the qubit case, where the bound $\mathcal{O}(1/n)$ can be  analytically obtained for $0\leqslant \alpha < 1$. 
	
	Let $V= Z$ and $H=X$,  where $X$ and $Z$ are the first and third Pauli matrix, respectively,  and take for simplicity $t=1$.
	Let 
\begin{equation}
U_n=\bigl(\rme^{-\rmi\frac{n^\alpha}{n} Z} \rme^{-\rmi\frac{1}{n}X}\bigr)^n, \qquad V_n=\rme^{-\rmi(n^\alpha Z+X)}.
\end{equation} 
We will  prove that
	\begin{equation}\label{eq:QubitAnalyticBound}
		U_n-V_n=\mathcal{O}\left(\frac{1}{n}\right).
	\end{equation}
For this purpose, first note that
	\begin{equation}
	\rme^{-\rmi\frac{n^\alpha}{n} Z}\rme^{-\rmi\frac{1}{n}X}=\left(\cos  \frac{n^\alpha}{n} \, I-\rmi\sin  \frac{n^\alpha}{n}\, Z\right)\left(\cos \frac{1}{n}\, I-\rmi\sin \frac{1}{n}\, X\right)
	=\rme^{-\rmi\theta_n \vec{u}_n\cdot \vec{\sigma}}
	\end{equation}
	where $\vec{\sigma}=(X,Y,Z)$,
	\begin{equation}
		\theta_n=\arccos\left(\cos \frac{n^\alpha}{n} \, \cos \frac{1}{n} \right),
	\end{equation}
	and
	\begin{equation}
		\vec{u}_n=\frac{1}{\sin \theta_n }\left(\cos \frac{n^\alpha}{n}\, \sin \frac{1}{n},\, \sin  \frac{n^\alpha}{n}\,\sin \frac{1}{n},\,\sin\frac{n^\alpha}{n}\,\cos\frac{1}{n}\right)
	\end{equation}
	is a unit vector, $\abs{\vec{u}_n}=1$. Thus we have
	\begin{equation}
		U_n=\rme^{-\rmi n\theta_n \vec{u}_n\cdot \vec{\sigma}}, \qquad
		V_n=\rme^{-\rmi \phi_n \vec{v}_n\cdot \vec{\sigma}},
	\end{equation}
	where 
	\begin{equation}
		\phi_n=\sqrt{n^{2\alpha}+1}, \qquad \vec{v}_n=\frac{1}{\phi_n}\left(1,0,n^\alpha \right).
	\end{equation}
	Therefore,
	\begin{align}\notag
		U_n-V_n&=\left(\rme^{-\rmi n\theta_n\vec{u}_n\cdot\vec{\sigma}}-\rme^{-\rmi\phi_n\vec{u}_n\cdot\sigma}\right)+\left(\rme^{-\rmi\phi_n\vec{u}_n\cdot\vec{\sigma}}-\rme^{-\rmi\phi_n\vec{v}_n\cdot\vec{\sigma}}\right)\\
		&=\rme^{-\rmi\phi_n \vec{u}_n\cdot \vec{\sigma}}\left(\rme^{\rmi(\phi_n-n\theta_n)\vec{u}_n\cdot\vec{\sigma}}-I\right)-\rmi\sin \phi_n\, (\vec{u}_n -\vec{v}_n)\cdot \vec{\sigma} .
				\label{eq:Un-Vn}
	\end{align}
	It follows that the distance between the two evolutions is controlled by the differences $\phi_n-n\theta_n$ and $\vec{u}_n-\vec{v}_n$, whose asymptotic is
	\begin{equation}\label{eq:AnaliticBound1}
		\phi_n-n\theta_n
		\sim \frac{1}{6} \frac{n^\alpha}{n^2}, 
		\qquad
		\vec{u}_n-\vec{v}_n \sim \left( -\frac{1}{3} \frac{n^\alpha}{n^2},\, \frac{1}{n} ,\, -\frac{1}{6}\frac{1}{n^2} \right),
	\end{equation}
as $n\to\infty$.
By plugging~\eqref{eq:AnaliticBound1} 
into Eq.~\eqref{eq:Un-Vn}, we finally get
\begin{equation}
U_n-V_n \sim -\rmi \frac{1}{n} \sin \phi_n\, Y,
\end{equation}
as $n\to\infty$, that is~\eqref{eq:QubitAnalyticBound}. Notice that the dominant term in the convergence error comes from the difference of the unit vectors, $\vec{u}_n-\vec{v}_n= \mathcal{O}(1/n)$, the phase difference being of smaller order, $\phi_n-n\theta_n = o (1/n)$.
	
	\section{Conclusions and outlook}

By providing a different scaling in the exponent of product evolutions, we have provided a bridge between periodically kicked systems, Trotter product formulas and strong-coupling limits. Our studies with numerical and analytical examples indicate a surprisingly good scaling of the error terms obtained. This paves the way to more efficient quantum control techniques:  while in the standard scaling Eq.~\eqref{standard} for bang-bang control the total time of the control pulses (in \figurename~\ref{fig:picfeynman} on the right the total base length of blue rectangles) grows as $n$, using shorter kicks (having the same strength) the same limit can be approached with a total duration of the control pulses scaling as  $n^\alpha$, at the price of a slower convergence rate.

Further work is needed in order to elucidate the $\mathcal{O}(1/n)$ behavior of the error~(\ref{eq:numerr}), numerically observed and discussed in Sec.~\ref{numan}.  The qubit's example discussed in Sec.~\ref{sec:qubit} might serve the purpose, and in particular the fact that the convergence rate comes from the error $\mathcal{O}(1/n)$ on the eigenprojections, the error on the eigenvalues being of smaller order---a fact that might be of general nature.

	\section*{Acknowledgments}
	We thank Kazuya Yuasa for discussions.
	PF, GG and SP are partially supported by Istituto Nazionale di Fisica Nucleare (INFN) through the project ``QUANTUM".
	PF and GG are partially supported by the Italian National Group of Mathematical Physics (GNFM-INdAM).

	\appendix
	
	\section{Proof of Theorem~\ref{thm:PulsedFormulation}}
	\label{sec:app1}
	This result has been proven in~\cite{ref:bernad} by using mean erdogic theorems. Furthermore, it can be obtained as a corollary of the analogous result, valid for general quantum operations, proven in~\cite{ref:unity2}. Here, however, by exploiting unitarity, we are able to give a simpler and direct proof with an explicit error bound, which is suitable to the generalisation we aim for in Theorem~\ref{thm:doublelimit}. (See \ref{sec:appinterm}.)

	First, noting that $\Uk=\Uk^{k}\Uk^{\dagger k-1}$, we can rewrite the product formula as
	\begin{align}\notag
	\Uk^{\dagger n}\bigl(\Uk \rme^{-\rmi \frac{t}{n}H}\bigr)^n
	&=\Uk^{\dagger n} \Uk \rme^{-\rmi \frac{t}{n}H}\cdots \Uk \rme^{-\rmi \frac{t}{n}H}\\\notag
	&=\Uk^{\dagger n-1} \rme^{-\rmi \frac{t}{n}H}\Uk^{n-1}\cdots \Uk^\dagger \rme^{-\rmi \frac{t}{n}H}\Uk \rme^{-\rmi \frac{t}{n}H}\\
	&=\rme^{-\rmi \frac{t}{n}H_{n-1}}\rme^{-\rmi \frac{t}{n}H_{n-2}}\cdots \rme^{-\rmi \frac{t}{n}H_0},
	\end{align}
	where we have defined
	\begin{equation}
	H_k:=\Uk^{\dagger k}H\Uk^k, \qquad (k=0,\dots,n-1).
	\end{equation}
	In the following we will need to expand the product of the exponentials in power series, and since in this expansion we will consider products among  different rotated Hamiltonians $H_k$ which do not commute with each other, it is convenient to introduce a notation for an ordered product. If $A=\{j_1,\dots, j_m\}$ is a set of indexes with $j_1\leqslant j_2 \leqslant \dots \leqslant j_m$, we define the ordered product as
	\begin{equation}
	\mathcal{T}\Big(\prod_{k\in A} H_k \Big):= H_{j_m}H_{j_{m-1}}\cdots H_{j_1} .
	\end{equation}
	The proof consists of two steps: first, we show that
	\begin{equation}\label{eq:firststep}
	\mathcal{T}\Big(\prod_{k=0}^{n-1}\rme^{-\rmi \frac{t}{n}H_k}\Big)-\rme^{-\rmi t\overline{H}_n}=\mathcal{O}\left(\frac{1}{n}\right),
	\end{equation}
	where
	\begin{equation}
	\overline{H}_n=\frac{1}{n}\sum_{k=0}^{n-1}H_k ;
	\end{equation}
	then, we show that
	\begin{equation}\label{eq:secondstep}
	\overline{H}_n=H_Z+\mathcal{O}\left(\frac{1}{n}\right).
	\end{equation}
	Once we have proven~\eqref{eq:firststep} and~\eqref{eq:secondstep}, we can combine them to obtain the statement of the theorem.
	
	In order to prove~\eqref{eq:firststep}, we proceed by expanding in series the exponentials:
	\begin{align}\label{eq:prodexp-expsum}\notag
	&\rme^{-\rmi  \frac{t}{n}H_{n-1}}\cdots  \rme^{-\rmi  \frac{t}{n}H_0} -\rme^{-\rmi  \frac{t}{n}(H_0+\dots+H_{n-1})} \\\notag
	&\qquad=\sum_{k_0,\dots,k_{n-1}}\left(-\frac{\rmi t} {n}\right)^{k_0}\cdots\left(-\frac{\rmi t} {n}\right)^{k_{n-1}}\frac{H_{n-1}^{k_{n-1}}}{k_{n-1}!}\cdots\frac{H_0^{k_0}}{k_0!} \\\notag 
	&\qquad\quad -\sum_{l=0}^\infty \frac{1}{l!}  \left(-\frac{\rmi t} {n}\right)^l(H_0+\cdots+H_{n-1})^l\\
	&\qquad=\sum_{l=0}^\infty \frac{1}{l!} \left(-\frac{\rmi t} {n}\right)^l R_l(n) ,
	\end{align}
	with $R_l(n)$ given by
	\begin{equation}
	R_l(n)= \sum_{k_0+\cdots+k_{n-1}=l}
	\binom{n}{k_0,\dots, k_{n-1}}\mathcal{T}\big(H_{n-1}^{k_{n-1}}\cdots H_0^{k_0}\big)
	-\left(H_0+\cdots +H_{n-1}\right)^l ,
	\end{equation}
	where
	\begin{equation}
	\binom{n}{k_0, \dots, k_{n-1}}=\frac{n!}{k_0!\cdots k_{n-1}!}
	\end{equation}
	is the multinomial coefficient. 
	Now we note that $\mathcal{T}(H_{i_1}\cdots H_{i_m})$ is invariant under permutation of the indexes:  inside the $\mathcal{T}$-ordering the rotated Hamiltonians commute with each other, and we can therefore use  the multinonial theorem to write the first term as
	\begin{equation}
	\sum_{k_0+\cdots+k_{n-1}=l}\binom{l}{k_0, \dots, k_{n-1}}\mathcal{T}(H_{n}^{k_{n}}\cdots H_0^{k_0})
	=\mathcal{T}\Big[\Big(\sum_{k=0}^{n-1}H_k\Big)^l\Big].
	\end{equation}
	Plugging this result into the expression of $R_l(n)$ we get
	\begin{equation} 
	R_l(n)=\mathcal{T}\Big[\Big(\sum_{k=0}^{n-1}H_k\Big)^l\Big]-\Big(\sum_{k=0}^{n-1}H_k\Big)^l
	=\sum_{k_1,\dots,k_l=0}^{n-1}[\mathcal{T}\left(H_{k_1}\cdots H_{k_l}\right)-H_{k_1}\cdots H_{k_l}] ,
	\label{eq:Rl(n)}
	\end{equation}
	for $l\geq 2$, while $R_0(n)=R_1(n)=0$.
	Since we are computing $l$ sums ranging over $n$ terms, in general we would expect this term to be $\mathcal{O}\left(n^l\right)$. However, using the spectral decomposition it is possible to obtain a better bound. Indeed, let~\eqref{eq:specdecUk}
	be the spectral decomposition of $\Uk$, where $\rme^{-\rmi \phi_\mu}\neq \rme^{-\rmi \phi_\nu}$ for $\mu\neq\nu$. Then
	\begin{equation}
	H_{k_1}\cdots H_{k_l}= \!\! \sum_{\mu_0,\dots,\mu_l=1}^m \!\! P_{\mu_0}H_{k_1}P_{\mu_1}\cdots P_{\mu_{l-1}}H_{k_l}P_{\mu_l} ,
	\end{equation}
	so that for each index $k_j$  we get
	\begin{equation}
	\sum_{k=0}^{n-1}P_{\mu}H_kP_{\nu} =\sum_{k=0}^{n-1}\rme^{-\rmi k(\phi_\nu-\phi_\mu)}P_\mu H P_\nu 
	= c_n(\phi_\nu-\phi_\mu) P_\mu H P_\nu,
	\label{eq:ksumoverPHkP}
	\end{equation}
	where 
	\begin{equation}\label{eq:cnphidef}
	c_n(\phi) =
	\left\lbrace
	\begin{aligned}
	&\frac{1-\rme^{-\rmi n \phi}}{1-\rme^{-\rmi \phi}} 
	& \phi \neq 0\\
	& n  &\phi=0
	\end{aligned}
	\right. .
	\end{equation}
	Notice that $c_n(\phi)$ is a continuous function, bounded by its value at $0$, namely $|c_n(\phi)|\leq c_n(0) = n$ and one gets for all $\mu\neq \nu$
	\begin{equation}
	|c_n(\phi_\mu-\phi_\nu)| 
	\leq C,
	\end{equation}
	with an $n$-independent bound
	\begin{equation}\label{eq:C}
	C:=
	\max_{\mu\neq\nu} \abs{\sin \left(\frac{\phi_\mu-\phi_\nu}{2}\right)}^{-1}.
	\end{equation}
	Then the dominant terms in the summation over $\mu_0,\dots,\mu_l$ are the diagonal ones,  corresponding to $\mu_0=\mu_1=\dots=\mu_l$, which are $\mathcal{O}(c_n(0)^l)=\mathcal{O}(n^l)$, but in such a case the two terms in square brackets of~\eqref{eq:Rl(n)} cancel out, since $P_\mu H_k P_\mu = P_\mu H P_\mu$ for all $\mu$ and $k$, whence
	\begin{equation}
	\mathcal{T}(P_{\mu}H_{k_1}P_{\mu}H_{k_2}P_{\mu}\cdots P_{\mu}H_{k_l}P_{\mu})=(P_\mu H P_\mu)^l
	=P_{\mu}H_{k_1}P_{\mu}H_{k_2}P_{\mu}\cdots P_{\mu}H_{k_l}P_{\mu}.
	\end{equation}
	Once we have understood that the highest order in equation~\eqref{eq:Rl(n)} is missing, we can claim that in general $R_l(n)$ will be $\mathcal{O}\left(n^{l-1}\right)$, since when the indexes $\mu_j$ are not all equal to each other, there is a $j$ such that $\mu_{j-1}\neq \mu_{j}$: we can perform the sum over $k_{j}$ using~\eqref{eq:ksumoverPHkP} which gives the $\mathcal{O}(1)$ term $c_n(\phi_{\mu_j}-\phi_{\mu_{j-1}})$, whose modulus is bounded by $C$. Then we can bound the remaining $l-1$ sums using the triangle inequality, so that	
	\begin{eqnarray}
\fl \qquad	\quad \Big\|\sum_{\bm{k}} P_{\mu_0} H_{k_1} P_{\mu_1} \cdots H_{k_l} P_{\mu_l}\Big\| 
	&\leq& \sum_{\bm{k}^{(j)}} \Big\| \sum_{k_j} P_{\mu_0} H_{k_1} P_{\mu_1} \cdots H_{k_l} P_{\mu_l}\Big\| 
	\nonumber\\
	&=& \sum_{\bm{k}^{(j)}} | c_n(\phi_{\mu_j}-\phi_{\mu_{j-1}})| \Norm{ P_{\mu_0} H_{k_1} P_{\mu_1} \cdots H_{k_l} P_{\mu_l}} 
	\nonumber\\
	&\leq& C \sum_{\bm{k}^{(j)}}  \Norm{ P_{\mu_0}} \Norm{H_{k_1}} \Norm{P_{\mu_1}} \cdots \Norm{H_{k_l}} \Norm{P_{\mu_l}} 
	\nonumber\\
	&=& C \sum_{\bm{k}^{(j)}}  \Norm{H}^l  
	= C n^{l-1} \Norm{H}^l ,
	\label{eq:bound1}
	\end{eqnarray}	
	where for convenience we set $\bm{k} =(k_1,k_2, \dots, k_l)$ and $\bm{k}^{(j)} =(k_1, \dots, k_{j-1}, k_{j+1}, \dots, k_l)$, and
	used $\Norm{P_\mu}=1$ for all $\mu$, and $\Norm{H_k}=\Norm{H}$ for all $k$.
	
	The ordered product can be bounded analogously by noting that the sum
	$\sum_{\bm{k}} \mathcal{T} (H_{k_1}\cdots H_{k_l} )$ is equal to the sum of $l!$ terms of the form
	\begin{equation}
	\sum_{\bm{k} \in \Delta_l} H_{k_1}\cdots H_{k_l} = \sum_{\bm{\mu}} \sum_{\bm{k} \in \Delta_l} P_{\mu_0}H_{k_1}P_{\mu_1}\cdots H_{k_l}P_{\mu_l},
	\end{equation}
	where the sum over $\bm{k}$ is restricted to 
	\begin{equation}
	\Delta_l = \{ \bm{k} = (k_1,\dots,k_l) \,:\, n-1\geq k_1 > k_2 > \cdots > k_l \geq 0  \}
	\end{equation}
	or  to similar ordered sets with some $>$ replaced by $\geq$.
	Consider a given $\bm{\mu}=(\mu_0,\dots, \mu_l)$, and as above assume that $\mu_{j-1}\neq \mu_{j}$ for some $j$. Then
	\begin{eqnarray}
\fl \qquad\quad 	\Big\|\sum_{\bm{k} \in \Delta_l } P_{\mu_0} H_{k_1} P_{\mu_1} \cdots H_{k_l} P_{\mu_l}\Big\| 
	&\leq& \sum_{\bm{k}^{(j)}\in\Delta_{l-1}} \! \Big\| \! \sum_{k_j = k_{j+1}+1}^{k_{j-1}-1} \! \! P_{\mu_0} H_{k_1} P_{\mu_1} \cdots H_{k_l} P_{\mu_l}\Big\| 
	\nonumber\\
	&\leq& \sum_{\bm{k}^{(j)}\in\Delta_{l-1}} \! C \Norm{ P_{\mu_0} H_{k_1} P_{\mu_1} \cdots H_{k_l} P_{\mu_l}} 
	\nonumber\\
	&\leq&   \sum_{\bm{k}^{(j)}\in\Delta_{l-1}} \!  C \Norm{H}^l 
	\leq    C \frac{n^{l-1}}{(l-1)!} \Norm{H}^l, 
	\end{eqnarray}	
	where in the last inequality we used 
	\begin{equation}
	\sum_{\bm{k}\in\Delta_{l-1}} 1 = \binom{n}{l-1} \leq \frac{n^{l-1}}{(l-1)!}.
	\end{equation}
	Thus we have
	\begin{equation}
	\Big\|\sum_{\bm{k}} \mathcal{T} \left(P_{\mu_0} H_{k_1} P_{\mu_1} \cdots H_{k_l} P_{\mu_l}\right) \Big\| \leq 
	C l  n^{l-1}\Norm{H}^l,
	\label{eq:bound2}
	\end{equation}
	so that, by gathering~\eqref{eq:bound2} and~\eqref{eq:bound1} and summing over $\bm{\mu}$ we finally obtain
	\begin{equation}
	\Norm{R_l(n)}  \leqslant \sum_{\bm{\mu}} C (l+1)  n^{l-1}\Norm{H}^l 
	\leqslant 2 C l  n^{l-1} m^{l+1} \Norm{H}^l,
	\end{equation}
	for $l\geq 1$.
	By plugging this result into equation~\eqref{eq:prodexp-expsum} we obtain
	\begin{align}\notag
	\big\| \rme^{-\rmi  \frac{t}{n}H_{n-1}}\cdots  \rme^{-\rmi  \frac{t}{n}H_0} -\rme^{-\rmi t\overline{H}_n} \big\| &\leqslant\sum_{l=1}^{\infty} \frac{1}{l!} \left(\frac{t}{n}\right)^l\Norm{R_l(n)}\\\notag
	&\leqslant \sum_{l=1}^{\infty} \frac{1}{l!} \left(\frac{t}{n}\right)^l 2C l n^{l-1}m^{l+1}\Norm{H}^l\\\notag
	&=\frac{2C t m^2  \Norm{H} }{n}\sum_{l=0}^{\infty}\frac{(tm\Norm{H})^l}{l!}\\
	&=\frac{2C t m^2  \Norm{H}\rme^{tm\Norm{H}}}{n} . 
	\end{align}
	Thus we have proved that
	\begin{equation}\label{eq:almostthere}
	\bigl\|\Uk^{\dagger n}\bigl(\Uk \rme^{-\rmi \frac{t}{n}H}\bigr)^n-\rme^{-\rmi t\overline{H}_n}\bigr\|\leq 
	\frac{2C t m^2  \Norm{H}\rme^{tm\Norm{H}}}{n}
	= \mathcal{O}\left(\frac{1}{n}\right) .
	\end{equation}

	Now, by using again the spectral decomposition~\eqref{eq:specdecUk} of $\Uk$ and  formula~\eqref{eq:ksumoverPHkP} we can see that the average Hamiltonian $\overline{H}_n$ tends to $H_Z$ as $n\rightarrow\infty$:
	\begin{equation}
	\overline{H}_n =\frac{1}{n}\sum_{k=0}^{n-1}\sum_{\mu,\nu=1}^m P_\mu H_k P_\nu
	=\sum_{\mu=1}^m P_\mu H P_\mu +\frac{1}{n}\sum_{\mu\neq\nu}
	c_n(\phi_\nu-\phi_\mu)
	P_\mu H P_\nu ,
	\end{equation}
whence
\begin{equation}
\Norm{ \overline{H}_n - H_Z} \leq \frac{1}{n}\sum_{\mu\neq\nu}
	|c_n(\phi_\nu-\phi_\mu)|\, \Norm{P_\mu H P_\nu} \leq \frac{C m^2 \Norm{H}}{n}.
\end{equation}
	As a consequence, we also get 
	\begin{equation}\label{eq:expHn=expHz}
	\bigl\| \rme^{-\rmi t\overline{H}_n} - \rme^{-\rmi tH_Z}\bigr\|
	\leq t \Norm{ \overline{H}_n - H_Z} \leq \frac{C t m^2 \Norm{H}}{n} = \mathcal{O}\left(\frac{1}{n}\right) 
	.
	\end{equation}
	Using~\eqref{eq:almostthere} and~\eqref{eq:expHn=expHz} we finally get 
\begin{equation}\label{eq:errorbound}
	\bigl\|\Uk^{\dagger n}\bigl(\Uk \rme^{-\rmi \frac{t}{n}H}\bigr)^n-\rme^{-\rmi t H_Z }\bigr\|\leq 
	\frac{C t m^2  \Norm{H}(1+2 \rme^{tm\Norm{H}})}{n}
	= \mathcal{O}\left(\frac{1}{n}\right) ,
	\end{equation}
and the theorem is proved.
		
	\section{Adiabatic Theorem}
	\label{sec:appadiab}
	
	The adiabatic theorem deals with the solution of the Schr\"{o}dinger equation
	\begin{equation}\label{eq:SchrodingerH(t)}
	i\frac{\rmd}{\rmd t}U(t)=\mathcal{H}(t)U(t), \qquad t\in[0,T],
	\end{equation}
	when the variation of the time-dependent Hamiltonian $\mathcal{H}(t)$ in the time interval $t\in[0,T]$ is very slow. There are several formulations of the adiabatic theorem~\cite{ref:avron}, differing both for the assumptions and the statement of the main theorem. Basically, the theorem states that an eigenstate of the initial Hamiltonian $\mathcal{H}(0)$ evolves into an eigenstate of the final Hamiltonian $\mathcal{H}(T)$ if the variation of $\mathcal{H}(t)$ is sufficiently slow. 
	
	Introducing a scaled time variable $s=t/T\in[0,1]$ and setting
	\begin{equation}
	H(s)=\mathcal{H}(sT),\qquad U_T(s)=U(sT),
	\end{equation}
	we can recast the Schr\"{o}dinger equation in the form
	\begin{equation}\label{eq:SchrodingerRescaled}
	i\frac{\rmd}{\rmd s}U_T(s)=TH(s)U_T(s), \qquad s\in[0,1],
	\end{equation}
	with $U_T(0)=I$.
	Then the limit of a very slow variation corresponds to the limit $T\rightarrow \infty$. By assuming that $H(s)$ is independent of $T$ we are requiring not only that the initial and final value of the Hamiltonian remain fixed while the time interval $[0,T]$ is stretched, but also that its ``shape'' is preserved. Kato proved the adiabatic theorem using a geometric approach defining an \textit{adiabatic evolution} $U(s)$ which rotates the eigenprojections of the initial Hamiltonian into the eigenprojections of the final Hamiltonian, and showing that the actual solution of equation~\eqref{eq:SchrodingerRescaled} approaches this adiabatic evolution when~$T\rightarrow\infty$.
	\begin{lemma}\label{lem:adiabatictransf}
		Let $P(s)$ be a twice continuously differentiable projection-valued function and define
		\begin{equation}
		A(s)=-i[\dot{P}(s),P(s)].
		\label{eq:Adef}
		\end{equation}
		Then the solution of the equation
		\begin{equation}\label{eq:adiabaticequation}
		\frac{\rmd}{\rmd s}U_{\mathrm{ad}}(s)=iA(s)U_{\mathrm{ad}}(s)
		\end{equation}
		with initial condition $U_{\mathrm{ad}}(0)=I$ satisfies the intertwining property
		\begin{equation}\label{eq:Intertwining}
		U_{\mathrm{ad}}(s)P(0)=P(s)U_{\mathrm{ad}}(s).
		\end{equation}
	\end{lemma}
	In the following we will refer to $U_{\mathrm{ad}}(s)$ as the adiabatic evolution
	\begin{thm}
		\label{thm:adiabatic}
		Let $H(s)$ be a family of Hermitian operators, $s\in[0,1]$ and let $P(s)$ be the instantaneous eigenprojection corresponding to the instantaneous eigenvalue $\lambda(s)$, namely
		\begin{equation}
		H(s)P(s)=\lambda(s)P(s).
		\end{equation}
		Assume that:
		\begin{itemize}
			\item [1)] $\lambda(s)$ is continuous,
			\item [2)] $P(s)$ is twice continuously differentiable.
		\end{itemize}
		Let $U_T(s)$  be the solution of equation~\eqref{eq:SchrodingerRescaled} with the initial condition $U_T(0)=I$. Then  
		\begin{equation}\label{eq:AdiabaticTheorem}
		\left[U_T(s)-\rme^{-iT\int_0^s \lambda(\sigma)\rmd \sigma} U_{\mathrm{ad}}(s)\right]P(0)=\mathcal{O}\left(\frac{1}{T}\right) ,
		\end{equation}
		as $T\rightarrow\infty$, where $U_{\mathrm{ad}}(s)$ is the adiabatic evolution in lemma~\ref{lem:adiabatictransf} 
	\end{thm}
	Using the intertwining property\eqref{eq:Intertwining} of the adiabatic transformation we can rewrite equation~\eqref{eq:AdiabaticTheorem} as
	\begin{equation}
	U_T(s)P(0)=\rme^{-iT\int_0^s\lambda(\sigma)\rmd\sigma}P(s)U(s)+\mathcal{O}\left(\frac{1}{T}\right) .
	\end{equation}
	We can see from this last expression the content of the adiabatic theorem: the evolution takes the range of $P(0)$ into the range of $P(s)$ plus a term which vanishes as $T\rightarrow\infty$.

	\section{Proof of Theorem~\ref{thm:ContFormulation}}\label{app:StrongCouplingProof}

	The proof makes  use of the adiabatic theorem reviewed in \ref{sec:appadiab}.
	See also~\cite{ref:unity2} for a generalisation to quantum semigroups.

	Consider the evolution in the interaction picture with respect to $H$:
	\begin{equation}
	\label{eq:UKI}
	U_K^I(t)=\rme^{\rmi  t H} \rme^{-\rmi  t (H +KV)}.
	\end{equation}
	It satisfies the Schr\"{o}dinger equation
	\begin{equation}\label{eq:interactionpictureequation}
	\rmi\frac{\rmd}{\rmd t} U_K^I (t)=KV^I(t)U_K^I(t),
	\end{equation}
	where $V^I(t)=\rme^{\rmi  t H }V\rme^{-\rmi  t H }$ is the control potential in the interaction picture. Its eigenprojections are $P_{\mu}(t)=\rme^{\rmi  t H}P_{\mu} \rme^{-\rmi  t H}$ and its eigenvalues are $\lambda_\mu$:
	\begin{equation}
	V^I(t)P_{\mu}(t)=\rme^{\rmi  t H} V \rme^{-\rmi  t H} \rme^{\rmi  t H}  P_{\mu} \rme^{-\rmi  t H}=\lambda_{\mu} P_{\mu}(t).
	\end{equation}
	Observe that
	\begin{equation}
	\frac{\rmd}{\rmd t}P_{\mu} (t)
	=\rmi [H, P_{\mu}(t)] .
	\end{equation}
	Notice that Eq.~\eqref{eq:interactionpictureequation} is of the same form of Eq.~\eqref{eq:SchrodingerRescaled}, 
	considered in the context of the adiabatic theorem, with the coupling constant $K$ playing the role of the parameter $T$ and the strong coupling limit corresponding to the adiabatic limit. 
	Both the eigenvalues and the eigenprojections are smooth functions of $t$. Therefore, from the adiabatic theorem~\ref{thm:adiabatic} we have
	\begin{equation}\label{eq:adiabaticlimitK}
	\left[ U_K^I(t)-
	\rme^{-\rmi t K \lambda_{\mu} }
	U_{\mathrm{ad}}(t)\right]P_{\mu}(0)= \mathcal{O}\left(\frac{1}{K}\right) .
	\end{equation} 
	The generator of the adiabatic transformation $U_{\mathrm{ad}}(t)$ given by~\eqref{eq:Adef} reads
	\begin{align}\notag
	A(t)&=-\rmi \bigl[\dot{P}_{\mu}(t), P_{\mu}(t)\bigr]
=\bigl[\left[ H ,P_{\mu}(t)\right],P_{\mu}(t)\bigr]=\rme^{\rmi  t H} \bigl[ [ H,P_{\mu}],P_{\mu}\bigr] \rme^{-\rmi  t H}
	=\rme^{\rmi tH}A\rme^{-\rmi tH}.
	\end{align}
	Therefore, $A(t)$ is the operator $A=\bigl[ [ H,P_{\mu}],P_{\mu}\bigr]$ in the $H$-interaction picture, and we get
	\begin{equation}
	U_{\mathrm{ad}}(t)=\rme^{\rmi  t H} \rme^{-\rmi  t (H -A)}.
	\end{equation}
	Thus~\eqref{eq:adiabaticlimitK} reads
	\begin{equation}
	\rme^{\rmi  t H} \bigl( \rme^{-\rmi  t (H +KV)}-\rme^{-\rmi  t K\lambda_{\mu}} \rme^{-\rmi  t (H -A)}\bigr) P_{\mu}(0)
	=\mathcal{O}\left(\frac{1}{K}\right) . \qquad 
	\end{equation}
	It remains to evaluate  $H-A$ explicitly,
	\begin{align}
	\notag H-A&= H -\bigl[ [H,P_{\mu}],P_{\mu}\bigr]
	=H-(H P_{\mu}-P_{\mu} H P_{\mu} -P_{\mu} H P_{\mu}+ P_{\mu} H)\\
	\notag&=H-(I -P_{\mu}) H P_{\mu}-P_{\mu} H (I - P_{\mu})\\
	&=P_{\mu} H P_{\mu}+(I - P_{\mu}) H (I - P_{\mu}) ,
	\end{align}
	so that by the preceding expression we obtain
	\begin{equation}
	\left( \rme^{-\rmi  t(H +KV)}-\rme^{-\rmi  t K\lambda_{\mu} } \rme^{-\rmi  t P_{\mu} H P_{\mu}}\right) P_{\mu}=\mathcal{O}\left(\frac{1}{K}\right) ,
	\end{equation}
	where we have taken into account that $P_{\mu}(0)=P_{\mu}$ and 
	\begin{equation}
	\rme^{-\rmi  t [P_{\mu} H P_{\mu}+(I - P_{\mu}) H (I - P_{\mu})]} P_{\mu}=\rme^{-\rmi  t P_{\mu} H P_{\mu} } P_{\mu}.
	\end{equation}
	Summing over $\mu=1,\dots,m$, using the completeness relation $\sum_{k}P_{\mu}=I $, and 
	\begin{align}
	\sum_{k=1}^m \rme^{-\rmi  t P_{\mu} H P_{\mu}} P_{\mu}
	=\sum_{k=1}^m \rme^{-\rmi  t H_Z} P_{\mu}
	=\rme^{-\rmi  t H_Z},
	\end{align}
	we  finally obtain
	\begin{equation}
	\rme^{-\rmi  t (H +KV)}- \rme^{-\rmi  t KV} \rme^{-\rmi  t H_Z}=\mathcal{O}\left(\frac{1}{K}\right) ,
	\end{equation}
	which completes the proof.

	\section{Proof of Theorem~\ref{thm:doublelimit} }
	\label{sec:appinterm}

	The proof is a corollary of the proof of Theorem~\ref{thm:PulsedFormulation} by setting
	\begin{equation}
	\Uk = \rme^{-\rmi \frac{t}{n}K_n V}, 
	\end{equation}
	whose spectral resolution is~\eqref{eq:specdecUk} with $\phi_\mu = t \lambda_\mu K_n/ n$. 
	Notice that, since by assumption $K_n/n\to 0$, for sufficiently large $n$ one has  $\max_{\mu,\nu} |\phi_\mu-\phi_\nu|\in (0,2\pi)$, 
	whence $\rme^{-\rmi \phi_\mu} \neq \rme^{-\rmi \phi_\nu}$ for all~$\mu\neq\nu$.
	
	As a consequence the bound $C$ in~\eqref{eq:C} reads
	\begin{equation}
	C =  \max_{\mu\neq\nu} \abs{\sin \frac{ K_n t(\lambda_\mu-\lambda_\nu)  }{2 n}  }^{-1}=\mathcal{O}\left(\frac{n}{K_n}\right) ,
	\end{equation}
	as $n\to\infty$.
	Accordingly, 
the bound~\eqref{eq:errorbound} become
\begin{equation}\label{eq:errorboundK}
\bigl\| \rme^{\rmi tK_n V}\bigl(\rme^{-\rmi \frac{t}{n}K_n V}\rme^{-\rmi \frac{t}{n}H}\bigr)^n -  \rme^{-\rmi tH_Z} \bigr\|\leq 
\frac{C t m^2  \Norm{H}(1+ 2 \rme^{tm\Norm{H}})}{n}
= \mathcal{O}\left(\frac{1}{K_n}\right) ,
\end{equation}
and we get~\eqref{eq:IntermediateLimit}.

	\section{Proof of Theorem~\ref{thm:GeneralizedProductFormula}}
	\label{app:GenTrotter}
	
	Let
	\begin{equation}
	A_n=-\rmi K_n V=K_n A, \qquad  B=-\rmi  {t} H, 
	\label{eq:AnB}
	\end{equation}
	and observe that  
	\begin{equation}
	U=\rme^{A_n/n}\rme^{B/n}, \qquad  W=\rme^{( {A_n}+B)/n}
	\end{equation} 
	are unitary operators. One gets
	\begin{align}
	\bigl\| \bigl(\rme^{A_n/n}\rme^{B/n}\bigr)^n - \rme^{A_n+B}\bigr\|&=\Norm{U^n-W^n}
	=\Bigl\| \sum_{k=0}^{n-1}U^k(U-W)W^{n-1-k} \Bigr\| \nonumber\\
	& \leqslant \sum_{k=0}^{n-1}\Norm{U^k(U-W)W^{n-1-k}} 
	=\sum_{k=0}^{n-1}\Norm{U-W} \nonumber\\
	& =n\Norm{U-W} .
	\end{align}
	By expanding the exponentials in series and gathering terms of the same order in $n$ we get
	\begin{align}\label{eq:GeneralizedTrotterFormulaDerivation}
	\notag\bigl\| \bigl(\rme^{A_n/n}\rme^{B/n}\bigr)^n - \rme^{A_n+B} \bigr\|&
	\leqslant n \bigl\| \rme^{A_n/n}\rme^{B/n}-\rme^{(A_n+B)/n} \bigr\| \\
	\notag &= n\Bigl\| \frac{[A_n,B]}{2n^2}+\sum_{l=3}^\infty\frac{C_l}{l!n^l} \Bigr\|
	=\Bigl\| \frac{[K_n A,B]}{2n}+\sum_{l=3}^\infty \frac{C_l}{l!n^{l-1}} \Bigr\|\\
	& \leqslant \frac{K_n \Norm{A}\Norm{B}}{n}+ \sum_{l=3}^\infty \frac{\|C_l\|}{l! n^{l-1}},
	\end{align}
	with
	\begin{align}
	 C_l=\sum_{k=0}^l \binom{l}{k}A_n^k B^{l-k}-(A_n+B)^l 
	=\sum_{k=1}^{l-1} K_n^k  \bigg[\binom{l}{k}A^k B^{l-k}-(\overbrace{A^k B^{l-k}+\dots}^{\binom{l}{k} \text{ permutations}}) \bigg].
	\end{align}
	Therefore, by setting $M=\max\{\Norm{A},\Norm{B}\}$, we  get
	\begin{align}
	\notag 
	\sum_{l=3}^\infty\frac{\Norm{C_l}}{l! n^{l-1}}
	&\leqslant \sum_{l=3}^\infty\frac{1}{l! n^{l-1}}\sum_{k=1}^{l-1} 2 K_n^k  \binom{l}{k}\Norm{A}^k\Norm{B}^{l-k}
	 \leqslant \sum_{l=3}^\infty \frac{1}{l!n^{l-1}} \sum_{k=1}^{l-1} 2 K_n^k  \binom{l}{k}M^l\\ 
	& = 2\sum_{l=3}^\infty \frac{M^l K_n}{l! n^{l-1}}\sum_{k=0}^{l-2}\binom{l}{k+1} K_n^{k}, 
	\end{align}
	so that, since 
	\begin{equation}
	\binom{l}{k+1} = \frac{l!}{(k+1)! (l-1-k)!} \leqslant \frac{l!}{k! (l-2-k)!},
	\end{equation}
	one gets
	\begin{align}
	\notag \sum_{l=3}^\infty \frac{\|C_l\|}{l! n^{l-1}} 
	& \leqslant 2\sum_{l=3}^\infty \frac{M^l K_n}{(l-2)! n^{l-1}}\sum_{k=0}^{l-2}\binom{l-2}{k} K_n^{k}
	= 2 \sum_{l=3}^\infty \frac{M^l  K_n  (K_n+1)^{l-2}}{(l-2)! n^{l-1}}\\ 
	&\leqslant \frac{2 M^3 K_n (K_n +1)}{n^2} \sum_{l=0}^\infty \frac{1}{l!}\left( \frac{M (K_n +1)}{n}\right)^l
	= \mathcal{O} \left(\frac{K_n (K_n +1)}{n^2}\right).
	\end{align}
	This means that the second term in Eq.~\eqref{eq:GeneralizedTrotterFormulaDerivation} can be neglected and we  finally  {obtain}
	\begin{equation}
	\bigl\| \bigl( \rme^{A_n/n}\rme^{B/n}\bigr )^n - \rme^{A_n+B} \bigr\| \leqslant \frac{K_n \|A\| \Norm{B}}{n}+o\left(\frac{K_n}{n}\right),
	\end{equation}
	whence, by definition~\eqref{eq:AnB},
	\begin{equation}\label{eq:GeneralizedProdFormula}
	\bigl(\rme^{-\rmi \frac{t}{n}V} \rme^{-\rmi \frac{t}{n}H}\bigr)^n - \rme^{-\rmi t(K_n V+H)}=\mathcal{O}\left(\frac{K_n}{n}\right), 
	\end{equation}
	as $n\to\infty$,
	which completes the proof.

\end{document}